%% file: main.tex
\documentclass[aip,amsmath,amssymb,preprint,jcp]{revtex4-1}

\draft 

\usepackage{graphicx} 
\usepackage{dcolumn}
\usepackage{bm} 

\include{article.preamble}

\usepackage[utf8]{inputenc}
\usepackage[T1]{fontenc}
\usepackage{mathptmx}
\usepackage{color}
\usepackage{csquotes}
\usepackage{siunitx}
\usepackage{mathtools}

\begin{document}

\title{Quasi-classical simulations of resonance Raman spectra based on path integral linearization}


\author{Hugo Bessone}
\email{hugo.bessone@sorbonne-paris-nord.fr}
\author{Rodolphe Vuilleumier}
\email{rodolphe.vuilleumier@ens.fr}
\affiliation{PASTEUR, Département de chimie, École normale supérieure, PSL University, Sorbonne Université, CNRS, 75005 Paris, France}
\author{Riccardo Spezia}
\email{riccardo.spezia@sorbonne-universite.fr}
\affiliation{Sorbonne Université, Laboratoire de Chimie Théorique, UMR 7616 CNRS, 4 Place Jussieu, 75005 Paris (France)}




\date{\today}

\begin{abstract} 
Based on a {linearization approximation} coupled with path integral formalism, we propose a method derived from the propagation of quasi-classical trajectories
to simulate resonance Raman spectra. 
This method is based on a ground state sampling followed by an ensemble of trajectories on the mean surface between the ground and excited states. 
The method was tested on three models and compared to quantum mechanics
solution based on a sum-over-states approach: harmonic and anharmonic oscillators and the HOCl molecule (hypochlorous acid). 
The method proposed is able to correctly characterize resonance Raman scattering and enhancement, including the description of overtones and combination bands. The absorption spectrum is obtained at the 
same time and the vibrational fine structure can be reproduced for long excited state relaxation times. The method can be applied also to dissociating 
excited states (as is the case for HOCl). 
\end{abstract}


\keywords{Resonant Raman Spectrum --- Molecular Dynamics --- Path Integral --- Linearization}

\maketitle



\section{Introduction}

Resonance Raman (RR) spectroscopy is a well-established tool to investigate molecular structures in different environments. 
For example, in biochemistry it is possible to identify vibrational signatures of specific chromophores removing the contribution
of the environment on the spectra\cite{carey_1978}. In fact, RR spectroscopy consists in irradiating the sample with an incident laser pulse whose energy corresponds (or is close) to
an electronic transition and then record the Raman spectra. The resulting Raman intensity of the absorbing molecules is tremendously enhanced and this makes 
the RR technique a powerful tool in analytical, physical and biological chemistry\cite{asher93}.
Moreover, for the purpose of species identification in complex media, RR not only offers selectivity through tuning of the incident light but it also enhances overtones and combination bands, leading to complex fingerprints even for simple species. This has been the case, for example, for the radical S$_3^{\bullet -}$, the species at the origin of the deep blue colour of ultramarine and lapis lazuli\cite{Chivers2013}. While it has only three vibrational bands, up to six overtones and combination bands are visible using RR\cite{Chivers1972,Lede2007}. Recently, S$_3^{\bullet -}$ has been identified in high pressure hydrothermal systems through the detection of an overtone using RR\cite{Pokrovski2011,Jacquemet2014,Pokrovski2015}.
The same principles of RR are at the basis of its use in material science and of surface enhanced Raman spectroscopy (SERS)\cite{campion98}.

The general theory of RR was given by Albrecht as a special case of Raman scattering\cite{Albrecht61} based on the time-independent Kramers-Heisenberg-Dirac (KHD) formula\cite{Kramers1925,Dirac1927}.
This approach is at the basis of calculation of RR spectra of molecules given their 
equilibrium geometry\cite{Mennucci07}. It can be combined with time-dependent formulation for the excited-state 
and both solvation and anharmonicity can be included\cite{Egidi2014,Baiardi2018}.


An alternative approach, based on a time-dependent reformulation of KHD expression, was developed by Heller and co-workers\cite{Lee79,Tannor82,Heller82}.
This development provides an expression of RR spectrum in terms of wave-packet nuclear dynamics in the short-time limit and it was at the basis, for example, of the 
method of Ben-Nun and Martinez, who used the ab-initio multiple spawning method from which RR spectrum of ethylene was successfully obtained\cite{BenNun99}.
Another semi-classical approach, based on Herman-Kluk propagator, was proposed by Voth and co-workers, obtaining the RR spectrum of I$_2$ in Xe at 230~K in agreement with experiments\cite{Ovchinnikov2001}.

When computing RR  to study chromophores
in complex  environments, like, e.g. carotenoids in different solvents, into an isolated protein and in the full biological
photo-system\cite{Ruban95,Kish2015,Macernis2015,Mendes2013,Mendes2013b,Lutz87,Koyama83},
the time-independent KHD formalism faces a challenge: one has to define a 
minimum energy structure and then obtain the associated normal modes. This can be problematic for the many flexible molecules, who are particularly interesting to investigate through vibrational spectroscopy. In proteins (and more in general in biological media), where geometries with minimum energy topology can be hard (albeit impossible) to be located. 
For example, it is 
known that an environment will allow distortion from planarity of carotenoids (a typical chromophore where RR studies provide much information) but the minimum energy searches fail for such configurations and theoretical spectra are 
often calculated only for planar minimum energy structures\cite{Kish2014,Kish2015,Macernis2015}. The time-dependent formalism offers a tempting possibility. However, the semi-classical methods developed so far can hardly be extended
to complex systems mainly due to high computational cost. Further developments and approximations are thus needed to apply this
approach to extended molecular systems.

In general, a practical and successful way of performing theoretical vibrational spectroscopy of complex molecular systems
is using molecular dynamics and spectral density methods. Examples can be found in biophysics\cite{Curutchet17}, 
ion chemistry\cite{galimberti2019}
or crystallography\cite{jahnigen2018}. 
Based on linear response theory, the Fourier transform of the appropriate time-correlation function can provide several spectroscopic signals. This approach can be used in the framework of molecular dynamics simulations, obtaining different
spectroscopic signals, like e.g.
IR\cite{gaigeot2015}, Raman\cite{Kaminiski2010}, vibration circular dichroism\cite{Scherrer2016}, vibrational sum frequency generation\cite{morita2002,sulpizi2013}. 
While nuclei generally evolve classically, the resulting intensities can be corrected using appropriate pre-factors to include quantum effects\cite{Valleau2012}. 
Recently, Mennucci and co-workers obtained approximated RR signals from biomolecular simulations\cite{Bondanza2020,Macaluso2020} by combining spectral density (obtained from autocorrelation function of excitation energies) with 
Huang-Rhys factors, giving a first order approximation of the
RR intensity of a discrete set 
of vibrational modes\cite{Page81,Sholz2011}. 

Additionally, the RR signal could be calculated as the auto-correlation of polarizability along a molecular trajectory through the so-called Placzek approximation, consisting in neglecting the vibrational state dependence of the transition energy from a vibrational state $i$ of the electronic ground state to a vibrational state $k$ of an electronic excited state: $\omega_{ki}\approx \frac{\Delta E_{el}}{\hbar}$, where $\Delta E_{el}$ is the electronic transition energy\cite{Walter2020}. Equivalently, the same result can be obtained from a classical, or short time, approximation to the Wigner transform of the polarizability operator\cite{Lee1983,Jensen2005}. The polarizability operator then depends only on the atomic positions and taking the derivative of the Placzek polarizability with respect to the position can provide the Raman intensity of fundamental transitions\cite{Walter2020,Jensen2005,Kane2010}. 
Recently, Resonance Raman spectra were thus obtained from DFT-based molecular dynamics simulations by calculating for each step the polarizability
tensor using real time time-dependent DFT (RT-TDDFT)\cite{Mattiat2021,Brehm2020}. This is a further development of a previous work from which the Raman spectrum was obtained\cite{Luber2014} from the
Placzek’s polarizability theory\cite{Lee83}.
RT-TDDFT was also used by other authors to obtain the Placzek's polarizability tensor and then RR spectra\cite{Jensen2005,Thomas2013}.

Here we present an extension of these works and we develop a correlation-function based approach that goes beyond the short time Placzek approximation, to obtain the signal from molecular dynamics simulations while still accounting for vibronic effects and describing overtones and combination bands. 
We start from the time-dependent formulation of KDH expression in the path-integral formalism and apply a forward-backward linearization of the path integral\cite{Sun1999,Poulsen2003,Shi2003,Shi2003b,Shi2003c,beutier2014}. Linearization has been shown to be a powerful framework to build semiclassical approximations to quantum correlation functions, mostly the linearized semi-classical initial value representation (LSC-IVR)\cite{Liu2006,Liu2007,Liu2008,Liu2011,Liu2015}. Our approach uses an extra linearization procedure for approximating the frequency dependent polarizability, which improves over the typical Placzek type approximation, in particular for the intensity of overtones and combination bands. 
This second linearization is closely related to linearization for non-adiabatic processes\cite{Sun1998,Bonella2005,Bonella2005b,Bonella2010}. It leads to a molecular dynamics scheme to compute RR spectra and also absorption spectra including vibronic effects.

The algorithm derived and proposed in the present work is applied to simple model systems for which it can be compared with
sum-over-states exact results. 
We first consider two-dimensional independent mode displaced harmonic oscillators (IMDHO)\cite{Walter2020} and then introduce non-harmonicity using a model due to Heller\cite{Heller82}. Finally, we consider the case of a dissociative electronic excited state through 
an analytical model of HOCl, proposed some years ago and used to model RR spectrum with a quantum dynamical treatment\cite{nanbu92,Offer96}.

 
%


\section{\label{sec:theory}Theory}

\subsection{Raman scattering and Absorption}

The Raman diffusion cross-section can be expressed from second-order perturbation theory and using the Born-Oppenheimer approximation as
\begin{equation}\label{eq:ramcross}
    I_{\text{Raman}}(\omega_s) \propto \omega_I\omega_s^3 \sum_i \rho(i) \sum_f \delta(\omega_I - \omega_s - \omega_{fi})\ \left\vert \alpha_{fi}(\omega_I) \right\vert^2
\end{equation}
where $i$ and $f$ denote the initial and final vibrational states, respectively, $\rho(i)$ is the probability for the system to be in $i$ state,
$\,\omega_I$ is the incident frequency, $\omega_s$ the scattered frequency, $\omega_{fi}$ is the energy difference between the initial  and final states and $\alpha_{fi}$ is the polarizability tensor.

The polarizability tensor, corresponding to a transition between vibrational states $f$ and $i$ of the electronic ground state $a$, can be expressed with the Kramers-Heisenberg-Dirac (KHD) equation\cite{Kramers1925,Dirac1927}:
\begin{align}\label{eq:khd}
  \begin{split}
    \alpha_{fi}(\omega_I)=\frac{1}{\hbar}\sum_b\sum_n&\left(\frac{\left<f\middle|\widehat{M_s}\middle|n\right>\left<n\middle|\widehat{M_I}\middle|i\right>}{\left(E_n-E_i\right)/\hbar-\omega_I-j\Gamma}\right.\\
     &\left.+\frac{\left<f\middle|\widehat{M_I}\middle|n\right>\left<n\middle|\widehat{M_s}\middle|i\right>}{\left(E_n-E_f\right)/\hbar+\omega_I+j\Gamma}\right)
  \end{split}
\end{align}

where $b$ denotes the electronic excited states and
$n$ the vibrational states belonging to $b$. 
$\widehat{M_s}$ and $\widehat{M_I}$ are the transition dipole operators between electronic states $a$ and $b$, for scattered ($s$) and incident ($I$) polarizations, respectively. 
Finally, $\Gamma$ is a phenomenological damping factor, which is related to the lifetime of state $b$ (note that here and hereafter we denote the imaginary unit as $j$ to avoid
confusion with initial vibrational state, $|i\rangle$).


The second term in Eq.~\ref{eq:khd} is the non resonant term $(NRT)$, conjugate to the first term with the opposite sign for $\omega_I$. Since in resonance Raman it is much smaller than the resonant term it is often omitted.


We first follow Jensen and Lee\cite{Jensen2005} to construct an approximation to the RR signal suitable to MD simulations by expressing Eq.~\ref{eq:ramcross} in the time-domain, by using the spectral representation of the Dirac distribution: 
\begin{equation}\label{eq:time2}
    \delta(\omega)=\frac{1}{2\pi}\int_{-\infty}^{+\infty}e^{-j\omega t}\,dt.
\end{equation}
Then, the RR intensity can  be written as
\begin{equation}
    I_\mathrm{Raman}(\omega_s)\propto \omega_I \omega_s^3 \sum_i \rho(i)
    \sum_f \frac{1}{2\pi} \int_{-\infty}^{+\infty}e^{-j(\omega_I-\omega_s) t}
    \alpha_{fi}^*(\omega_I) e^{j\frac{E_f}{\hbar} t}
    \alpha_{fi}(\omega_I)e^{-j\frac{E_i}{\hbar} t}
    \,dt.
\end{equation}
This can be transformed in a quantum correlation function by introducing a polarizability operator, $\widehat{\mathcal{P}}(\omega_I)$, through 
\begin{equation}
    \alpha_{fi}=\left<f\middle|\widehat{\mathcal{P}}(\omega_I)+\widehat{\mathcal{P}}^{NRT}(\omega_I)\middle|i\right>,
\end{equation}
with each element given by the KHD expression
\begin{align}
    \langle f\vert\widehat{\mathcal{P}}(\omega_I)\vert i \rangle=&
    \frac{1}{\hbar}\sum_b \sum_n \frac{\left<f\middle|\widehat{M_s}\middle|n\right>\left<n\middle|\widehat{M_I}\middle|i\right>}{\left(E_n-E_i\right)/\hbar-\omega_I-j\Gamma}\\ 
        \langle f\vert\widehat{\mathcal{P}}^{NRT}(\omega_I)\vert i \rangle=&
    \frac{1}{\hbar}\sum_b \sum_n
    \frac{\left<f\middle|\widehat{M_I}\middle|n\right>\left<n\middle|\widehat{M_s}\middle|i\right>}{\left(E_n-E_f\right)/\hbar+\omega_I+j\Gamma},
\end{align}
$\widehat{\mathcal{P}}^{NRT}(\omega_I)$ denoting the non-resonant term.
The RR signal is then expressed in the time domain by

\begin{widetext}

\begin{equation}
\begin{split}
    I_{\text{Raman}}(\omega_s) \propto \omega_I\omega_s^3 
    \int_{-\infty}^{+\infty} e^{-j(\omega_I-\omega_s)t} 
    \left\langle 
    \left( \widehat{\mathcal{P}}^{\dagger}(\omega_I)+\widehat{\mathcal{P}}^{NRT\dagger}(\omega_I) \right)
    e^{j\frac{\widehat{H_a}}{\hbar}t} 
    \left( \widehat{\mathcal{P}}(\omega_I)+\widehat{\mathcal{P}}^{NRT}(\omega_I) \right)
     e^{-j\frac{\widehat{H_a}}{\hbar}t} \right\rangle_\rho dt,
\end{split}
\end{equation}
\end{widetext}
where $\widehat{H_a}$ is the Hamiltonian operator for the system 
on state $a$, 
$\langle \cdot \rangle_\rho$ denotes an average over the density operator $\widehat{\rho}$ and $\widehat{A}^\dagger$ the adjoint of operator $\widehat{A}$. Neglecting the non-resonant term leads to the simpler expression:
\begin{equation}\label{eq:ramcrossTD}
\begin{split}
    I_{\text{Raman}}(\omega_s) & \propto \omega_I\omega_s^3 
    \int_{-\infty}^{+\infty} e^{-j(\omega_I-\omega_s)t} 
    \left\langle 
     \widehat{\mathcal{P}}^{\dagger}(\omega_I)
    e^{j\frac{\widehat{H_a}}{\hbar}t} 
    \widehat{\mathcal{P}}(\omega_I) 
     e^{-j\frac{\widehat{H_a}}{\hbar}t} \right\rangle_\rho dt \\
     & \propto \omega_I\omega_s^3 
    \int_{-\infty}^{+\infty} e^{-j(\omega_I-\omega_s)t} 
    \mathrm{Tr}\left( \widehat{\rho}
     \widehat{\mathcal{P}}^{\dagger}(\omega_I)
    e^{j\frac{\widehat{H_a}}{\hbar}t} 
    \widehat{\mathcal{P}}(\omega_I) 
     e^{-j\frac{\widehat{H_a}}{\hbar}t} \right) dt,
\end{split}
\end{equation}
The RR signal is now expressed as the Fourier transform of a quantum correlation function of the frequency dependent polarizability operator $\widehat{\mathcal{P}}(\omega_I)$ with dynamics on the electronic state $a$. 

Semiclassical approaches to the calculations of such correlation functions start by expressing it in terms of Wigner transforms of operators. For
a generic operator $\widehat{O}$, we can write the Wigner transform as:
\begin{align}
\begin{split}
O_W[x_0,p_0]&=\int_{-\infty}^{+\infty}\left<x_0-\frac{\Delta x_0}{2}\middle|\widehat{O}\middle|x+\frac{\Delta x_0}{2}\right>e^{+\frac{j}{\hbar}p_0\Delta x_0}\,d\Delta x_0 \\
&=\int_{-\infty}^{+\infty}\left<x_0+\frac{\Delta x_0}{2}\middle|\widehat{O}\middle|x_0-\frac{\Delta x_0}{2}\right>e^{-\frac{j}{\hbar}p_0\Delta x_0}\,d\Delta x_0.
\end{split}
\end{align}
We  quickly review such semiclassical approach for the specific case of RR. The quantum correlation function can be rewritten exactly as 
\begin{equation}\label{eq:wigner}
    I_\mathrm{Raman}(\omega_s)\propto \omega_I \omega_s^3 
    \int_{-\infty}^{+\infty} e^{-j(\omega_I-\omega_s)t} 
    \iint_{-\infty}^{+\infty}\!\!\!\!dx_0dp_0\left( \widehat{\rho}\widehat{\mathcal{P}}^{\dagger}(\omega_I)\right)_W[x_0,p_0] \left(\left(\widehat{\mathcal{P}}(\omega_I)\right)(t)\right)_W[x_0,p_0]\; dt,
\end{equation}
by using the following property of Wigner transforms:
\begin{equation}
    \mathrm{Tr}\gpar{\widehat{A}\widehat{B}}=\frac{1}{2\pi\hbar}\int A_W[x,p]B_W[x,p]\,dxdp.
\end{equation}

In Eq.~\eqref{eq:wigner}, we have the Wigner transform of the polarizability operator at time $t$ which is:
\begin{equation}
\left(\left(\widehat{\mathcal{P}}(\omega_I)\right)(t)\right)_W[x_0,p_0]=
\left(e^{j\frac{\widehat{H_a}}{\hbar}t} 
    \widehat{\mathcal{P}}(\omega_I) 
     e^{-j\frac{\widehat{H_a}}{\hbar}t}\right)_W[x_0,p_0]
\end{equation}
Several approaches, like initial value representation, have employed the linearization of the forward and backward time propagation 
to justify the semiclassical approximation\cite{Sun1998,Sun1999,Makri2002,Shi2003,Shi2003b,Poulsen2003,Liu2006,Liu2007,Liu2007b,Liu2008,Liu2011,Liu2015}:
\begin{equation}
    I_\mathrm{Raman}(\omega_s)\propto \omega_I \omega_s^3 
    \int_{-\infty}^{+\infty} dt e^{-j(\omega_I-\omega_s)t}
    \iint_{-\infty}^{+\infty}\!\!\!\!dx_0dp_0\left( \widehat{\rho}\widehat{\mathcal{P}}^{\dagger}(\omega_I)\right)_W[x_0,p_0] \left(\widehat{\mathcal{P}}(\omega_I)\right)_W[x_t,p_t]\
\end{equation}
where $(x_t,p_t)$ are the position and momenta at time $t$ along a classical trajectory initiated at time $t=0$ at $(x_0,p_0)$.

The  Wigner transform of
 $\widehat{\rho} \widehat{\mathcal{P}}^{\dagger}(\omega_I) $ represents 
 a challenge. Several works have tackled in which one has a product
 of density operator with position or momentum operators\cite{Liu2006,beutier2014,Bose2019}, however here the 
 polarizability operator is highly non-linear in $\widehat{x}$ and $\widehat{p}$, which gives rise to enhanced overtones and combination bands in RR spectra.
 Using the properties of Wigner transforms detailed in Appendix~\ref{app:wignerprod}, it is
 possible to obtain a simpler formulation of the Raman intensity:

\begin{equation}\label{eq:ramonce}
\begin{split}
    I_\mathrm{Raman}(\omega_s)\propto\frac{2}{1+e^{-\beta \omega}}
    \omega_I \omega_s^3
    \int_{-\infty}^{+\infty}\!\!\!\!dt\,e^{-j\omega t}\iint_{-\infty}^{+\infty}\!\!\!\!dx_0dp_0\left(\widehat{\rho}\right)_W[x_0,p_0]&\left(\widehat{\mathcal{P}}^{\dagger}(\omega_I)\right)_W[x_0,p_0]\\
    &\times\left(\widehat{\mathcal{P}}(\omega_I)\right)_W[x_t,p_t]
\end{split}
\end{equation}

Note that this expression coincides, at high temperature, with the classical approximation to the polarizablity-polarizability correlation
while the prefactor ensures that Stokes and anti-Stokes lines have the expected ratio of intensities at all temperatures.





\subsection{Linearization approach for the frequency dependent polarizability}\label{sec:linearpol}

To proceed further we need to express the Wigner transform of the frequency dependent polarization and here we propose a second linearization step.

As suggested by Heller\cite{Heller82}, the polarizability operator can be written in the time domain instead as a sum over vibrational states $\vert n\rangle$ in the electronic excited state $b$. In fact, using the well-known relation 
\begin{equation}\label{eq:time1}
    \frac{1}{\Omega}=j\int_{0}^{+\infty}e^{-j\Omega \tau}\,d\tau\quad ;\quad\text{Im}\{\Omega\}<0
\end{equation}
we obtain
\begin{equation}
    \langle f\vert\widehat{\mathcal{P}}(\omega_I)\vert i \rangle=
    \frac{j}{\hbar}\sum_b \sum_n \int_0^{+\infty}
    e^{-\Gamma \tau}e^{j\omega_I \tau}
    \left<f\middle|\widehat{M_s}\middle|n\right>e^{-\frac{j}{\hbar}E_n \tau}\left<n\middle|\widehat{M_I}\middle|i\right>e^{\frac{j}{\hbar}E_i \tau}
    \,d\tau.
\end{equation}
The resonant contribution to the polarizability can then be written as the element of the operator:
\begin{equation}\label{eq:polop}
    \widehat{\mathcal{P}}(\omega_I)=\frac{j}{\hbar}\sum_b\int_{0}^{+\infty}e^{-\Gamma\tau}e^{j\omega_I\tau}\,d\tau\,\widehat{M_s}\,e^{-\frac{j}{\hbar}\widehat{H_b}\tau}\,\widehat{M_I}\,e^{\frac{j}{\hbar}\widehat{H_a}\tau}
\end{equation}
where we discard again the non-resonant term.

The main difficulty for a semi-classical treatment of the polarizability operator is the propagation over a time $\tau$ backward on the ground state $a$, $e^{\frac{j}{\hbar}\widehat{H_a}\tau}$, and forward on the excited state $b$, $e^{-\frac{j}{\hbar}\widehat{H_b}\tau}$. Here, we apply linearization to a path integral formulation of these propagators and
we simplify the discussion for the case of only one excited state $b$. Note that it can be generalized to several excited states such that the
following discussion does not lose in generality. Final results generalized for more than one excited state are reported in Appendix~\ref{sec:generalb}.


We can first write the Wigner transform of the polarizability operator 
as:

\begin{align}
  \begin{split}\label{eq:pol2app}
    \gpar{\op{\mathcal{P}}(\omega_I)}_W[x,p]&=\intg{e^{\frac{j}{\hbar}p\Delta x}}{\Delta x}{-\infty}{+\infty}\frac{j}{\hbar} \intg{e^{-\Gamma\tau}e^{j\omega_I \tau}}{\tau}{0}{+\infty}\\
    &\times\braket{x-\frac{\Delta x}{2}}{\widehat{M_s}e^{-\frac{j}{\hbar}\op{H_b}\tau}\widehat{M_I}e^{\frac{j}{\hbar}\op{H_a}\tau}}{x+\frac{\Delta x}{2}}
  \end{split}
\end{align}

By applying the linearization procedure detailed in Appendix~\ref{sec:LPI} to the Wigner transform of the polarizability operator introduced above, and using the compact notation

\begin{align}\label{eq:varphi}
  \begin{split}
    \exp\gcro{-\frac{j}{2\hbar}\frac{\tau}{P+1}\sum_{k=0}^P
    \begin{pmatrix*}[l]
    V_b\gpar{\overline{x_k}}-V_a\gpar{\overline{x_k}}\\
    +V_b\gpar{\overline{x_{k+1}}}-V_a\gpar{\overline{x_{k+1}}}\\
    \end{pmatrix*}}&\approx\botacc{\exp\gcro{-\frac{j}{\hbar}\intg{\gpar{V_b(x_u)-V_a(x_u)}}{u}{0}{\tau}}}{\varphi_{\tau,fwd}^{b-a}(x,p)}\\
    &\approx\botacc{\exp\gcro{-\frac{j}{\hbar}\intg{\gpar{V_b(x_u)-V_a(x_u)}}{u}{-\tau}{0}}}{\varphi_{\tau,bwd}^{b-a}(x,p)}
  \end{split}
\end{align}


where a phase is accumulated over a backward trajectory propagated over the average surface, $V_m$, starting from $(x,p)$, we obtain:
\begin{equation}\label{eq:wigpola}
    \gpar{\op{\mathcal{P}}(\omega_I)}_W[x,p]=\frac{j}{\hbar} \intg{e^{-\Gamma \tau}e^{j\omega_I \tau}}{\tau}{0}{+\infty} M_s(x)M_I\gpar{x_{-\tau}}\varphi^{b-a}_{\tau,bwd}\gpar{x,p}
\end{equation}

and for the Wigner transform of the adjoint of the polarizability operator:

\begin{align}\label{eq:mainLPI}
  \begin{split}
    \gpar{\op{\mathcal{P}}^{\dagger}(\omega_I)}_W[x,p]&=-\frac{j}{\hbar} \intg{e^{-\Gamma \tau}e^{-j\omega_I \tau}}{\tau}{0}{+\infty} M_s(x)M_I\gpar{x_{-\tau}}\varphi^{a-b}_{\tau,bwd}\gpar{x,p}\\
    &=\gpar{\op{\mathcal{P}}(\omega_I)}_W^{\star}[x,p]
  \end{split}
\end{align}

Such trajectory on the mean surface of the ground and excited states is typical in non-adiabatic dynamics\cite{Bonella2005,Bonella2005b}. It is also found when deriving surface-hopping from mixed quantum-classical Liouville dynamics\cite{Kapral1999,Nielsen2000}, consistent with the fact that mixed quantum-classical Liouville dynamics can be itself derived through linearization\cite{Bonella2010}.

Introducing this result for the polarizability operator into the RR signal intensity, we get the final expression:
\begin{align}
  \begin{split}
    I_{Raman}(\omega_s)&=K\frac{2}{1+e^{-\beta \omega}}\omega_I\omega_s^3\frac{1}{2\pi}\frac{1}{\hbar^2}\intg{e^{-j\omega_s t}}{t}{-\infty}{+\infty}\intgd{\frac{\widehat{\rho}_W\gcro{x_0,p_0}}{2\pi\hbar}}{x_0}{p_0}{}\\
    &\times M_sM_I\intg{e^{-\Gamma \tau}e^{-j\omega_I\tau}\varphi^{a-b}_{\tau,bwd}(x_0,p_0)}{\tau}{0}{+\infty}\\
    &\times  M_sM_I\intg{e^{-\Gamma \tau'}e^{j\omega_I\tau'}\varphi^{b-a}_{\tau',bwd}(x_{t},p_{t})}{\tau'}{0}{+\infty}
  \end{split}
\end{align}

\subsection{Absorption}

The absorption spectrum is obtained simply from the equilibrium average
of the polarizability operator
\begin{equation}
    \mean{\op{\mathcal{P}}(\omega_I)}=\frac{j}{\hbar} \intg{e^{-\Gamma t}e^{j\omega_It}}{t}{0}{+\infty}\intg{\braket{x}{\widehat{\rho}\widehat{M_s}e^{-\frac{j}{\hbar}t\op{H_b}}\widehat{M_I}e^{\frac{j}{\hbar}t\op{H_a}}}{x}}{x}{-\infty}{+\infty}
\end{equation}

Introducing the linearized path-integral (LPI) result as in the previous section for the Wigner transform of the polarizability operator we obtain the LPI expression for absorption:
\begin{align}
  \begin{split}
    \mean{\op{\mathcal{P}}(\omega_I)}&=\frac{j}{\hbar}M_sM_I\intg{e^{-\Gamma t}e^{j\omega_It}}{t}{0}{+\infty}\intgd{}{x_0}{p_0}{}\\
    &\times\frac{\widehat{\rho}_W\gcro{x_0,p_0}}{2\pi\hbar}\varphi_{t,bwd}^{b-a}(x_0,p_0)
  \end{split}
\end{align}

This amounts to sampling initial conditions $(x_0,p_0)$ according to the Wigner transform of the density operator at finite temperature and for each initial condition performing a short molecular dynamics trajectory on the mean potential $\frac{1}{2}(V_a+V_b)$ to obtain the oscillating phase that is then Fourier transformed.

The linearization of the potentials around the average path is exact if both the ground state and excited states are harmonic potentials with the same curvature, that is for the case of displaced harmonic potentials. In that case, Eq.~\eqref{eq:mainLPI} is exact and so is then the LPI result for the Wigner transform of the polarizability operator. As no further approximation is needed to express the absorption from it, in this harmonic limit the absorption in the LPI approach is also exact. This will be illustrated below where we will show that LPI reproduces a virtually exact sum-over-states approach and thus the vibronic structure of the absorption spectrum.

\section{\label{sec:num}Numerical details}
\subsection{LPI-MD implementation}

From the expression of the signal, to compute the RR spectrum we first generate a set of initial conditions $x_0,p_0$ from the Wigner distribution. To do so  we have employed the Local Gaussian Approximation from Liu and Miller\cite{Liu2007}. It consists in a path-integral Monte-Carlo simulation to generate the initial positions, while momenta are generated randomly from a Gaussian distribution derived from the local curvature of the potential energy surface. The path-integral calculations were performed with 128 beads and at each step 16 consecutive beads were resampled using staging\cite{Sprik85}. 500 steps were used for equilibration and then initial conditions were extracted every 50 steps.

\begin{figure*}[htbp]\centering
\includegraphics[width=0.8\linewidth]{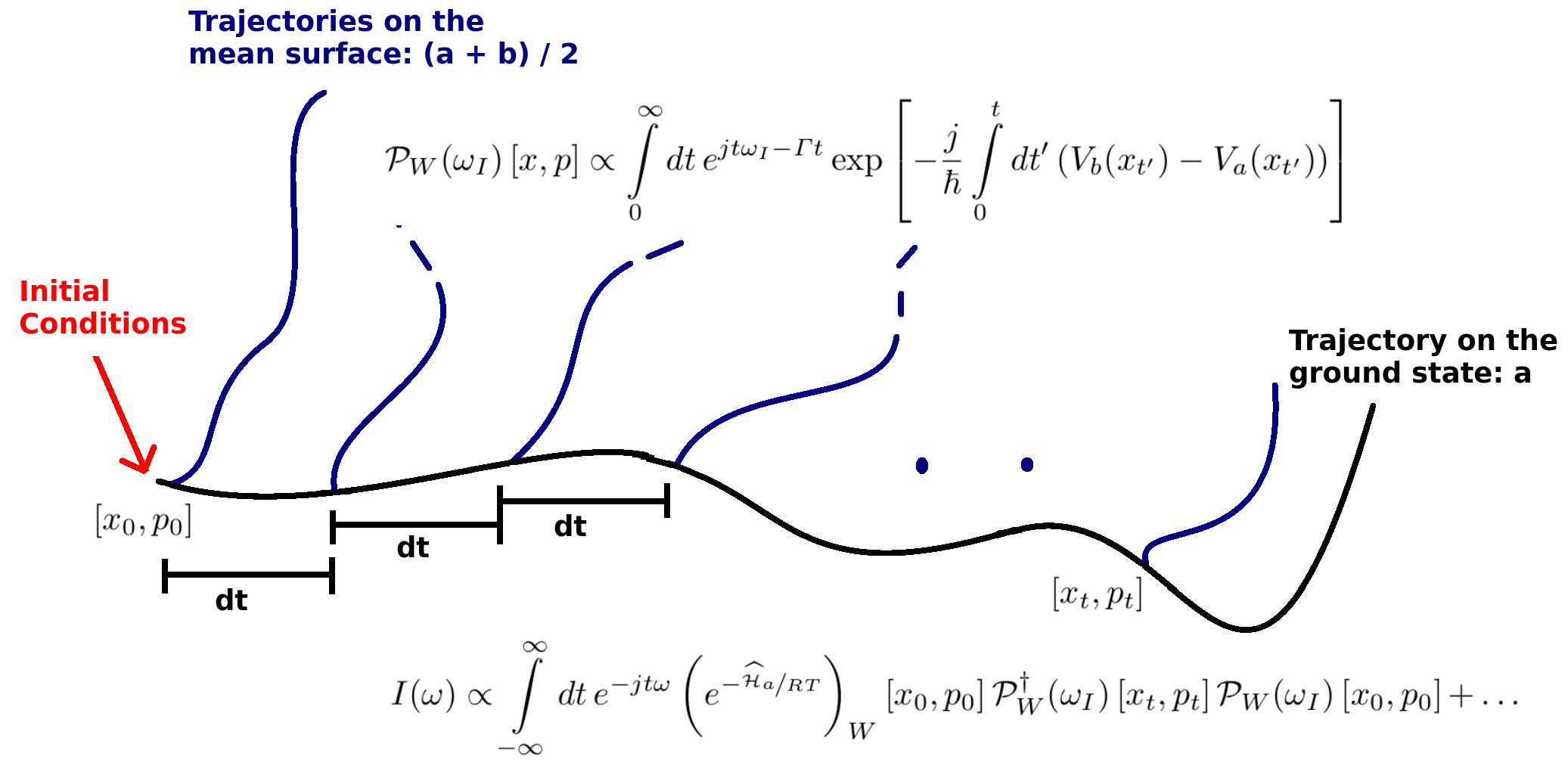}
\caption{\label{fig:res}Schematic representation of how we calculate Raman scattering from trajectory simulations.}
\end{figure*}

Figure~\ref{fig:res} visually summarizes the procedure to obtain
RR spectra from LPI-MD approach. From each initial configuration $x_0,p_0$, a trajectory $x_t,p_t$ is generated on the electronic ground state with a Velocity-Verlet algorithm (black line in Figure~\ref{fig:res}). Then at regular intervals, the frequency dependent polarizability is computed by running backward short trajectories on the mean ground and excited state surfaces, with initial configuration $x_t,p_t$, also using a Velocity-Verlet algorithm (purple lines in Figure~\ref{fig:res}). The Fourier transform of the phase factor, including the damping factor, along this backward trajectory gives the frequency dependent polarizability as a function of the incident frequency.  The auto-correlation function of the frequency dependent polarizability is then computed by correlating that quantity at time $t$ along the ground state trajectory with that quantity at time $t=0$. Note that we do not use a  \enquote{sliding} average over the ground state trajectory, since the sampling density is not conserved by the classical dynamics.

The Fourier transform of the autocorrelation function of the frequency dependent polarizability is then calculated for each incident frequency to obtain the RR spectrum as a function of both the incident and scattered frequencies.
As for the absorption spectra, one trajectory on the mean surface, starting from the configuration $x_0,p_0$, is enough to average the frequency dependent polarizability over the initial configurations.


The calculation of a RR spectrum is thus characterized by: (i) the number of initial configurations considered, (ii) the length $\Delta T$ of the trajectories on the ground state, (iii) the sampling interval $\delta T$ on this trajectory and (iv) the length $T_{aux}$ of the auxiliary trajectories on the mean surface. In the following, we will test the convergence over the number of initial configurations while the other parameters are dictated by the desired precision in the incident and scattered frequency. More specifically, the length of the trajectories on the ground state is related to the lowest accessible scattered frequency while the sampling interval to the highest desired frequency. Similarly, the length of the auxiliary trajectories conditions the accuracy of the RR signal as a function of the incident frequency. These parameters are given below for all simulations using the three models.

Finally, to make the figures easier to read, the Raman spectra peaks plotted with respect to the scattered frequency have all been widened. This has been done thanks to a convolution by a Cauchy–Lorentz distribution of 30~cm$^{-1}$ width.
Furthermore, when showing RR spectrum as a function of the scattered frequency, an integral over the incident frequency is performed and vice versa. In each case the integral of the incident and scattered frequencies correspond to integrating over a vibronic peak or over a vibrational band, respectively.


\subsection{Sum-over-states (SoS) implementation}

The reference SoS calculations were performed using a product discrete variable representation (DVR) of the 2-dimensional wavefunctions\cite{DVR_1982}. We have employed a Gauss-Hermite quadrature, expanding the wavefunctions with Hermite polynomials as eigenstates of an harmonic potential centered around a mid-point between the equilibrium positions in the ground and excited states. The curvature of that harmonic potential was adjusted to converge the vibrational wavefunction of both ground and excited states. The weights and quadrature positions were obtained by diagonalization of the position operator matrix, which was constructed recursively\cite{Golub1969,DVR_2002}. Typically, we have used $N_{\mathrm{dim}}=40$ basis functions in each direction, thus a total of $N_\mathrm{basis}=1600$ basis functions (convergence were tested with $N_{\mathrm{dim}}=50$ basis functions in each direction). To ensure convergence of the potential energy using Gauss-Hermite quadrature, we have used $N_\mathrm{points}=2N_\mathrm{dim}-1$ quadrature points in each direction. 



\section{\label{sec:2HO}IMDHO Model}

\subsection{Description}

We first apply the linearization approach to a 2-dimensional system for which both the ground and the excited state potential energy surfaces are modeled by harmonic oscillators (independent-mode displaced harmonic oscillators, IMDHO, model). 
We consider here the simplest case where the harmonic frequencies and modes are identical in both states, which corresponds to neglecting the Duschinsky rotation. In that case, the linearization step is exact. 
Non-harmonic models will be considered in the next sections.

The two potential energy surfaces are given by
\begin{align}
    V_\mathrm{gs}(x,y)&=\frac{1}{2}m\omega_x^2 x^2 + \frac{1}{2}m\omega_y^2 y^2 \label{eq:imdho1}\\
    V_\mathrm{es}(x,y)&=\frac{1}{2}m\omega_x^2 (x-x_e)^2 + \frac{1}{2}m\omega_y^2 (y-y_e)^2 + V_{01}, \label{eq:imdho2}
\end{align}
where $V_\mathrm{gs}$ and $V_\mathrm{es}$ are the ground-state and excited state energies as functions of the two degrees of freedom, $x$ and $y$,
that have the same mass $m$, $\omega_x$ and $\omega_y$ are the pulsations
of $x$ and $y$, respectively, being the same at ground and excited state, $V_{01}$ is the energy shift between ground and excited state and $x_e$ and $y_e$ are the shifts of the equilibrium position at the excited state for $x$ and $y$, respectively.
The values used for the different parameters are listed in Table~\ref{tab:00}. Unless otherwise stated, we have used a damping factor $\Gamma=0.02$~fs$^{-1}$, a reasonable value according to Jensen et al.\cite{Jensen2005}.



\begin{table}
\caption{\label{tab:00}Parameters used the IMDHO model as in Eqs.~\ref{eq:imdho1} and~\ref{eq:imdho2}. }
\begin{tabular}{c | c c}
\textbf{Parameter}       & \textbf{Value} & \textbf{Units} \\
\hline
$x_e$                   & 8.0 & pm \\
$y_e$                   & 12.0 & pm \\
$\omega_x$              & 2123 & cm$^{-1}$ \\
$\omega_y$              & 1592 & cm$^{-1}$ \\
$V_{01}$                & 10 & eV \\
$m$                     & 6.75 & g mol$^{-1}$ \\ 
\end{tabular}
\end{table}


\subsection{Absorption}

\begin{figure}[htbp]\centering
\includegraphics[width=\linewidth]{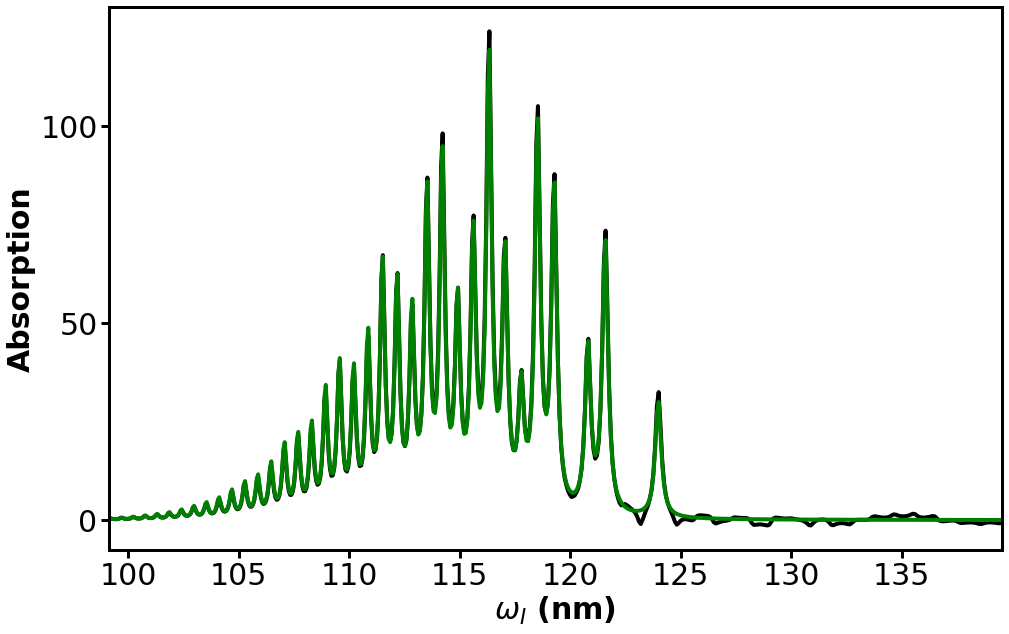}
\caption{\label{fig:01}Absorption spectra obtained from LPI-MD simulations at $300$ K for IMDHO model with $\Gamma =\nolinebreak 0.02$~fs$^{-1}$, 20~000 initial configurations and mean state trajectories each 300~fs long (black curve), compared with SoS results (green curve)}
\end{figure}

We have argued that LPI should be exact to describe the absorption spectrum for the IMDHO model. Figure~\ref{fig:01} displays a comparison between converged LPI results and SoS reference showing that LPI is in excellent agreement with the SoS result. 

\begin{figure}[htbp]\centering
\includegraphics[width=\linewidth]{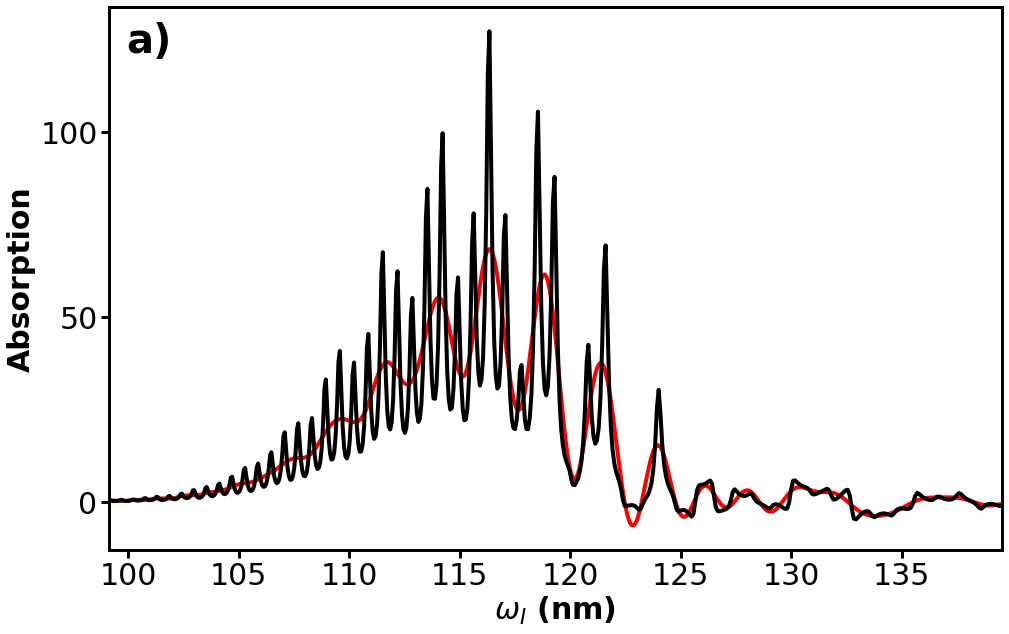}
\includegraphics[width=\linewidth]{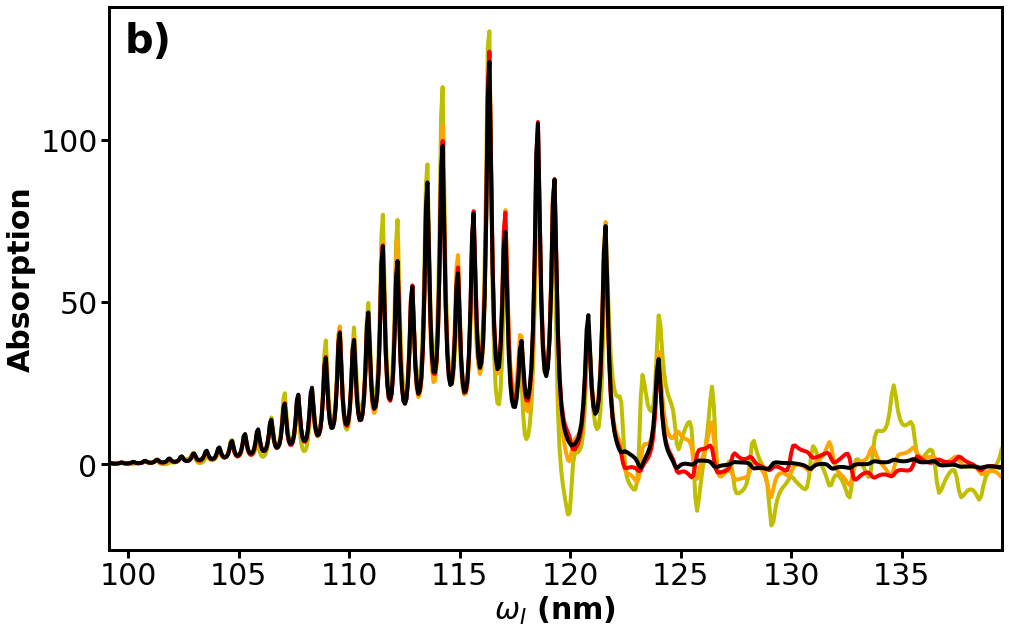}
\caption{\label{fig:03}Absorption spectra for IMDHO model obtained from LPI-MD simulations. Panel a): $30$ fs (red) or $300$ fs (black) lengths for the mean surface trajectories. 
Panel b): 100 (yellow), 500 (orange), 2000 (red) or 20~000 (black) initial configurations.}
\end{figure}

We will now briefly discuss how the LPI absorption spectrum depends on the different parameters of the semi-classical dynamics. First, we have varied the length of the trajectory on the mean state, using either 30~fs or 300~fs long mean trajectories. Figure~\ref{fig:03}.a shows that when the mean state trajectories are too short, the vibronic structure disappears in agreement with a lower resolution of the spectra according to the Nyquist-Shannon sampling theorem.

Another important parameter is the number of initial configurations used to sample the equilibrium Wigner density, which is then the number of mean surface trajectories. The absorption spectrum for different numbers of initial conditions is thus shown in Figure~\ref{fig:03}.b. While the spectrum at low wavelengths converges quickly with the number of initial conditions, this is not so at larger wavelengths where about 20~000 trajectories are needed. It appears that above 125~nm (see Figure~\ref{fig:01}), where the absorption spectrum is very small, the LPI signal results from a compensation of positive and negative random numbers. It is remarkable however that a positive signal is recovered only when enough statistics is acquired. This explains the small discrepancies between LPI-MD and SoS signals for $\omega_I >$~125~nm, which
will disappear in the limit of an infinit number of initial configurations.




\begin{figure}[htbp]\centering
\includegraphics[width=\linewidth]{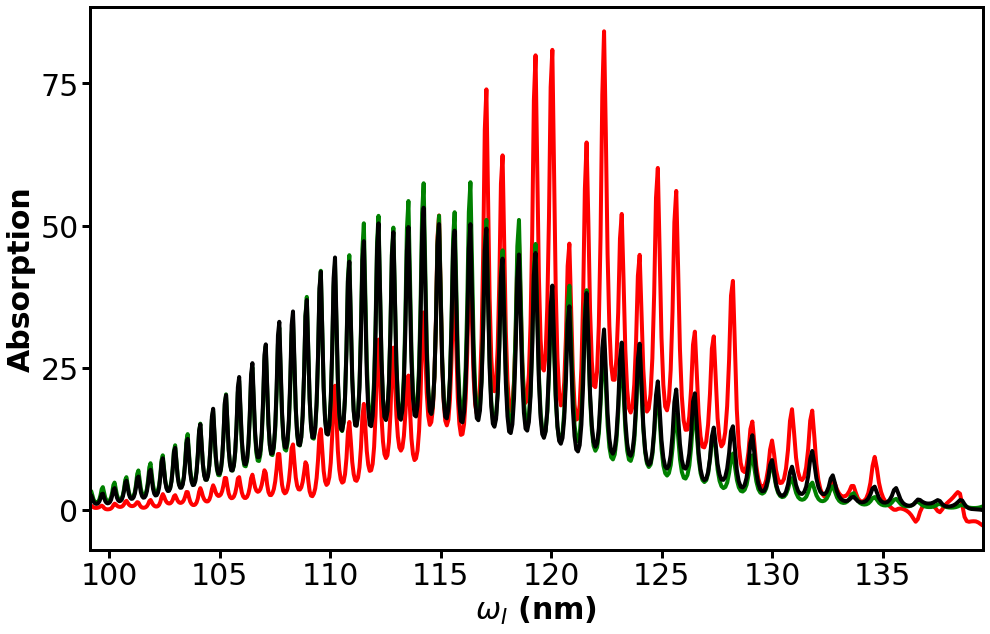}
\caption{\label{fig:04}Absorption spectra at $3000$ K for IMDHO model with $2000$ (red curve) or 20~000 (black curve) initial configurations compared with a \enquote{sum-over-states} approach (green curve)} 
\end{figure}

When increasing the temperature, from 300~K to 3000~K, the number of ground state trajectories
should also increase to obtain agreement with reference SoS results, as reported in Figure~\ref{fig:04}. This is probably due to the broader range of accessible configurations at this higher temperature.

In the following we will use for the resonance Raman 
spectra of the IMDHO model the simulation parameters listed in 
Table~\ref{tab:02}. By selecting the incident wavelength below 120~nm, we are able to use a lower number of mean surface trajectories, equal to 2000.

\begin{table}
\caption{\label{tab:02}Parameters used for the propagating the ground state and mean surface trajectories for IMDHO model.}
\begin{tabular}{l | c}
Parameter & Value \\
\hline
Timestep & 0.1  fs \\
\#~Ground state trajectories & 2000 \\
\#~Mean state trajectories & 10~000 \\
Length of ground state trajectories & 3000~fs \\
Length of mean state trajectories & 300~fs \\
Damping factor ($\Gamma$) & 0.02~fs$^{-1}$ \\
Scattered resolution & 0.01 rad.fs$^{-1}$
\end{tabular}
\end{table}



\begin{figure}[htbp]\centering
\includegraphics[width=\linewidth]{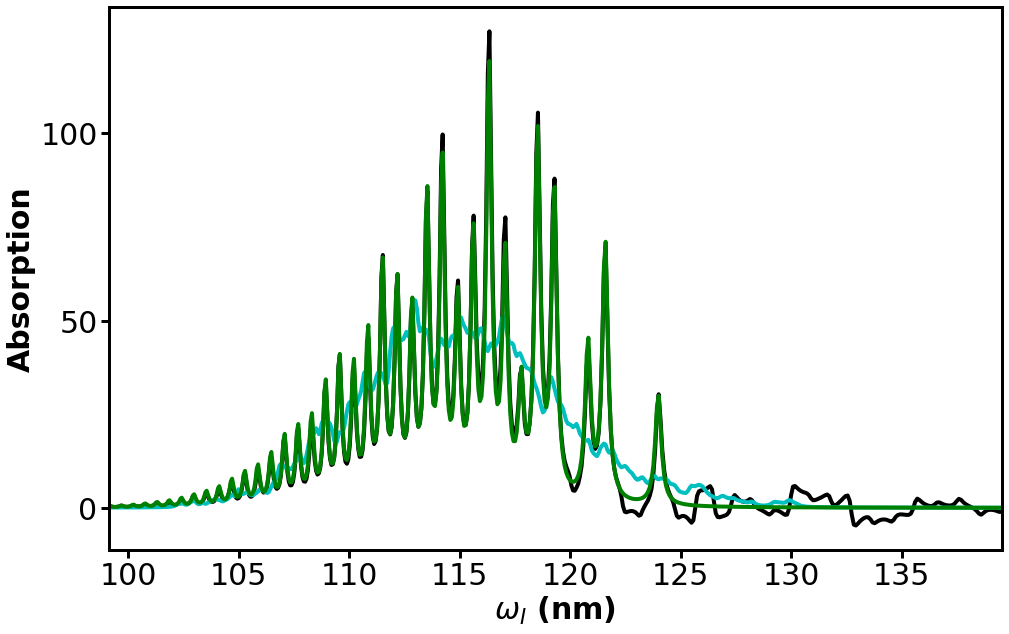}
\caption{\label{fig:35}
Absorption spectra obtained from LPI-MD simulations at $300$ K for IMDHO model with $\Gamma =\nolinebreak 0.02$~fs$^{-1}$,  (black curve), compared with SoS results (green curve) and Placzek type polarizability (cyan curve) approaches.}
\end{figure}

Finally, we compare in Figure~\ref{fig:35} our LPI-MD results with those using the Placzek approximation. LPI appears a clear improvement with respect to the Placzek approximation to reproduce the vibronic structure. Placzek approximation however correctly captures the overall shape of the spectrum, as if the damping factor was effectively larger.

\subsection{Resonance Raman}


We now focus on the RR spectrum of the IMDHO system. While in this ideal situation adsorption is exact within the linearization approximation, indicating that the correct expectation value of the frequency dependent polarizability operator is correctly obtained, further approximations are involved for the determination of the RR spectrum, since the time-dependent correlation function of the polarizability operator is obtained through classical trajectories with initial conditions sampled from the equilibrium Wigner density. 
Formally, for the calculation of the RR spetrum of the IMDHO model the only approximation is considering the Wigner transform of a product of operators as a product of their Wigner transforms. Note that for the absorption there is no approximation.  In this sub-section, we will thus investigate the effect of this approximation on the RR spectrum.

Two incident frequencies are used for investigating RR spectra: one at about 116~nm, corresponding to a \enquote{resonance} condition (it covers the highest intensity peak) and one at about 108~nm, which 
gives an example of \enquote{near-resonance} spectrum. 

\begin{figure}[htbp]\centering
\includegraphics[width=\linewidth]{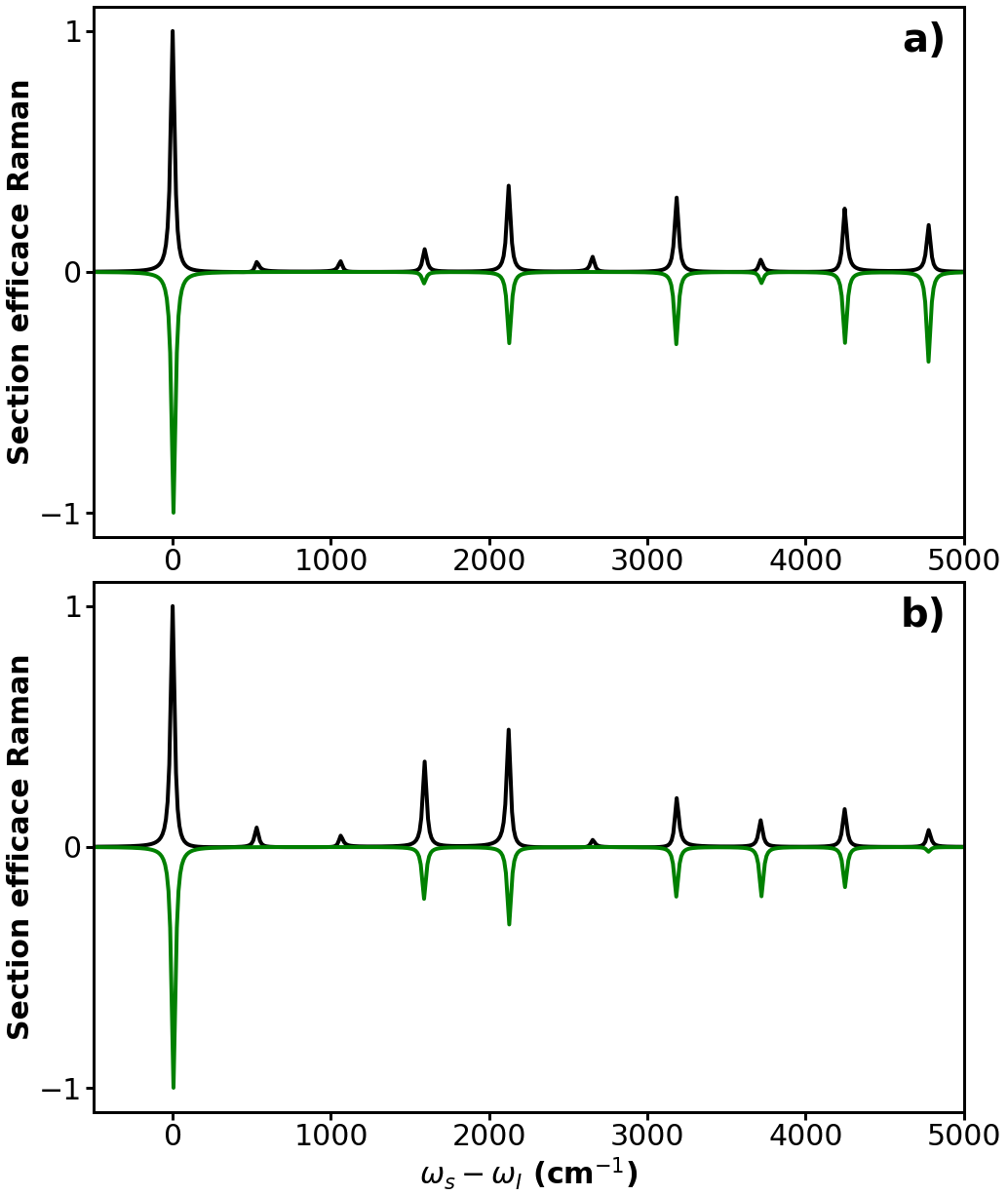}
\caption{\label{fig:10}Resonance Raman spectra of the IMDHO system: (a) \enquote{resonance}  and (b)  \enquote{near resonance} scattered frequencies. In both cases the signal is integrated over the whole frequency region. In black we report the LPI-MD results and in green the SoS ones. 
The intensities are normalized at the Rayleigh peak.}
\end{figure}

Figure~\ref{fig:10}.a shows the resonance Raman spectrum of the IMDHO system for an incident frequency in the \enquote{resonance} region using the LPI-MD approach and compared to the sum-over-states reference results. Both spectra are normalized to the Rayleigh peak and only the Stokes region is shown. 
The anti-Stokes peaks 
can be obtained from the well-known relation:
\begin{equation}
    \frac{I_{\text{Stokes}}}{I_{\text{anti-Stokes}}}=e^{2\beta\Delta E}
\end{equation}

where $\Delta E$~=~$\hbar(\omega_s - \omega_I)$.

Many peaks are observed, not only at the fundamental frequencies of the two modes, $\omega_x$ and $\omega_y$, but also for overtones and combination bands. These two frequencies are commensurate with and all peaks are thus multiples of $\Delta \omega=\omega_x-\omega_y=531~\text{cm}^{-1}$. The presence of these bands while the dynamics is harmonic is due to the highly non-linear character of the frequency dependent polarizability operator. The respective intensities of the multiple peaks are well reproduced up to about 5000~cm$^{-1}$. LPI-MD however seems to give rise to spurious peaks at 531~cm$^{-1}$, 1062~cm$^{-1}$ and 2655~cm$^{-1}$. These combination bands have in comparison very low 
intensities in the SoS reference. They may arise from coherence not captured by the product of Wigner transforms although they are small  and the global agreement between LPI-MD and SoS reference is satisfactory. This agreement  extends to an incident frequency in the low wavelength region of the absorption spectrum, as shown in Figure~\ref{fig:10}.b. 


\begin{figure}[htbp]\centering
\includegraphics[width=\linewidth]{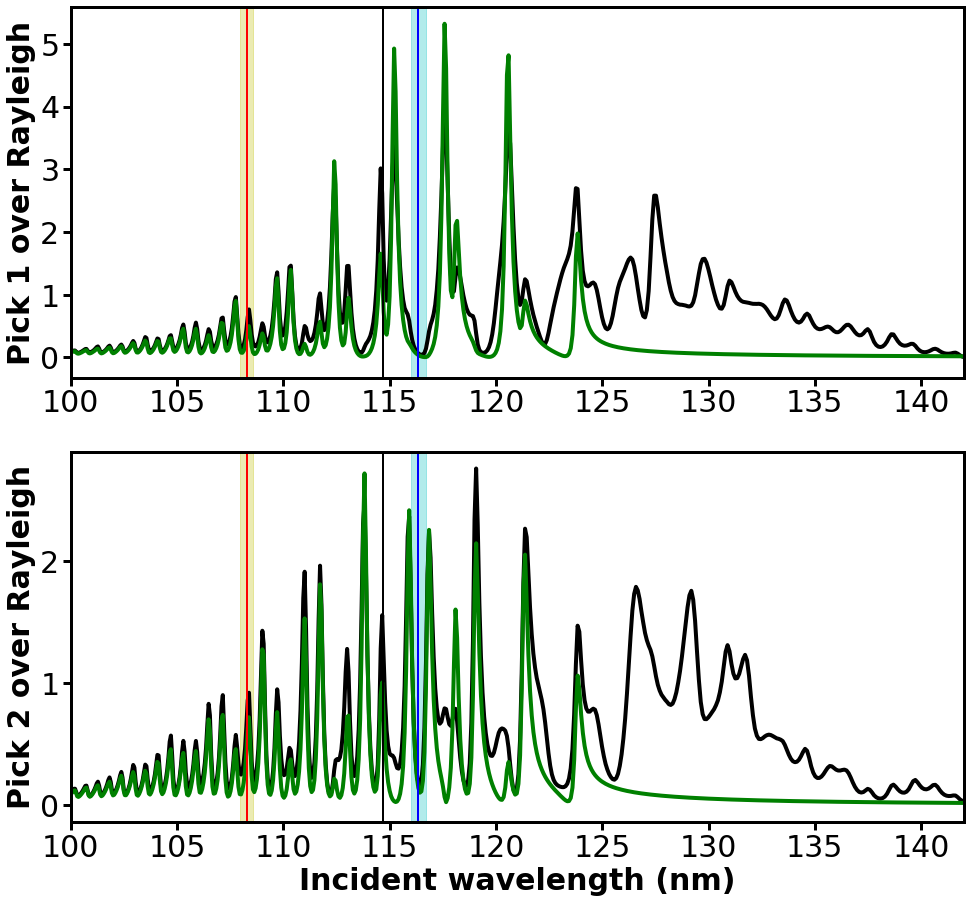}
\caption{\label{fig:07} Resonance Raman spectra of the IMDHO model 
with respect to the incident frequency divided by the spectra for the Rayleigh pick obtained with the LPI-MD approach (black curve), compared with the sum-over-states approach (green curve), for different scattered frequencies: the first peaks for both vibrational modes, with highlights on relevant pulsations (blue and red) and Franck-Condon frequency (black vertical line).}
\end{figure}

\begin{figure}[htbp]\centering
\includegraphics[width=\linewidth]{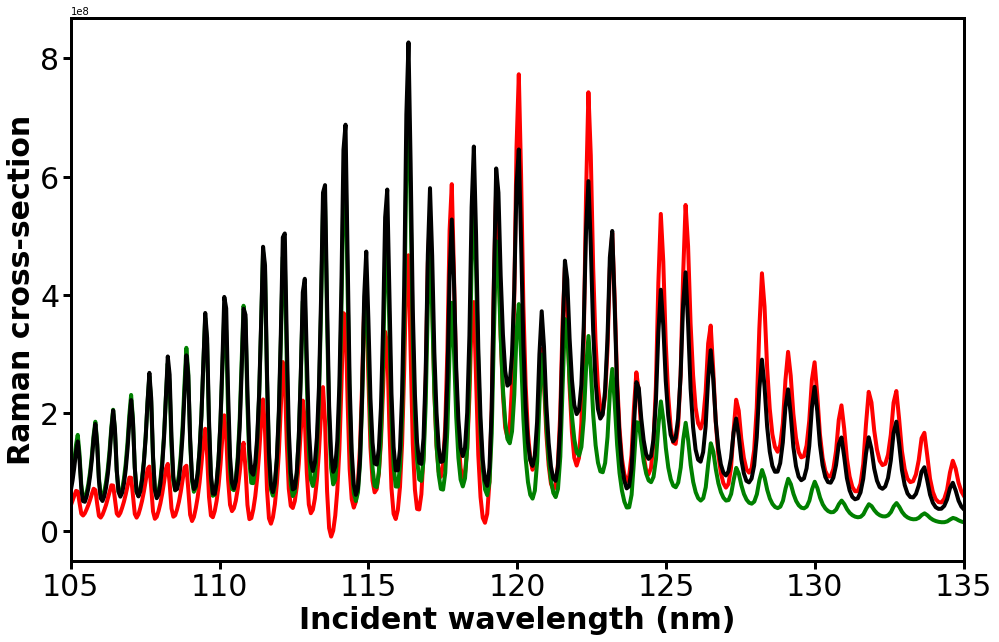}
\caption{\label{fig:18} Resonance Raman spectra for the IMDHO model at $3000$ K with respect to the incident frequency for the Rayleigh peak, using $2000$ (red curve) or $20~000$ (black curve) ground state trajectories, compared with the sum-over-states result (green curve).}
\end{figure}

A more stringent test is to examine the intensity of RR peaks $I(\omega_s-\omega_I)$ as a function of the incident frequency, $\omega_I$. 
To this end, we have integrated the intensity of the scattered peaks around both fundamental frequencies $\omega_s-\omega_I=\omega_x$ and $\omega_s-\omega_I=\omega_y$, as well as around the Rayleigh peak ($\omega_s-\omega_I=0$) for reference. The integration range around each peak is $\pm 100$~cm$^{-1}$.
The intensity of these two fundamental peaks normalized to the Rayleigh peak  is then shown in Figure~\ref{fig:07}.  Similarly to the absorption spectrum, the peak intensities are well reproduced in a broad region, including vibronic effect. However,  a discrepancy appears at higher wavelengths. This is a region with small absorbance where we have found that the signal arises from destructive interferences from different initial positions.  As we increase the temperature and the dynamics becomes more classical, this discrepancy decreases as it can be seen in Figure~\ref{fig:18}. Note that the larger range of initial conditions at high temperatures necessitates a higher number of sampled initial configurations (set to 20~000 in Figure~\ref{fig:18}).

\begin{figure}[htbp]\centering
\includegraphics[width=\linewidth]{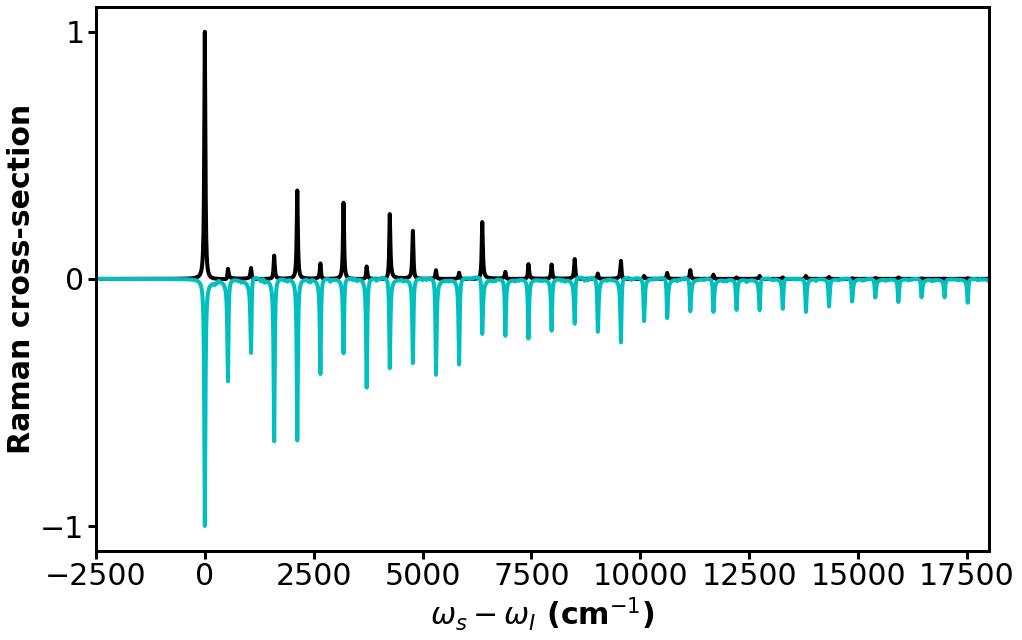}
\caption{\label{fig:36}Resonance Raman spectra of the IMDHO model with
respect a \enquote{near resonance} scattered frequency (the signal is integrated over the whole region), compared with results from the Placzek type polarizability (cyan curve). 
The intensities are scaled to have the same values for the Rayleigh peak.}
\end{figure}

\begin{figure}[htbp]\centering
\includegraphics[width=0.9\linewidth]{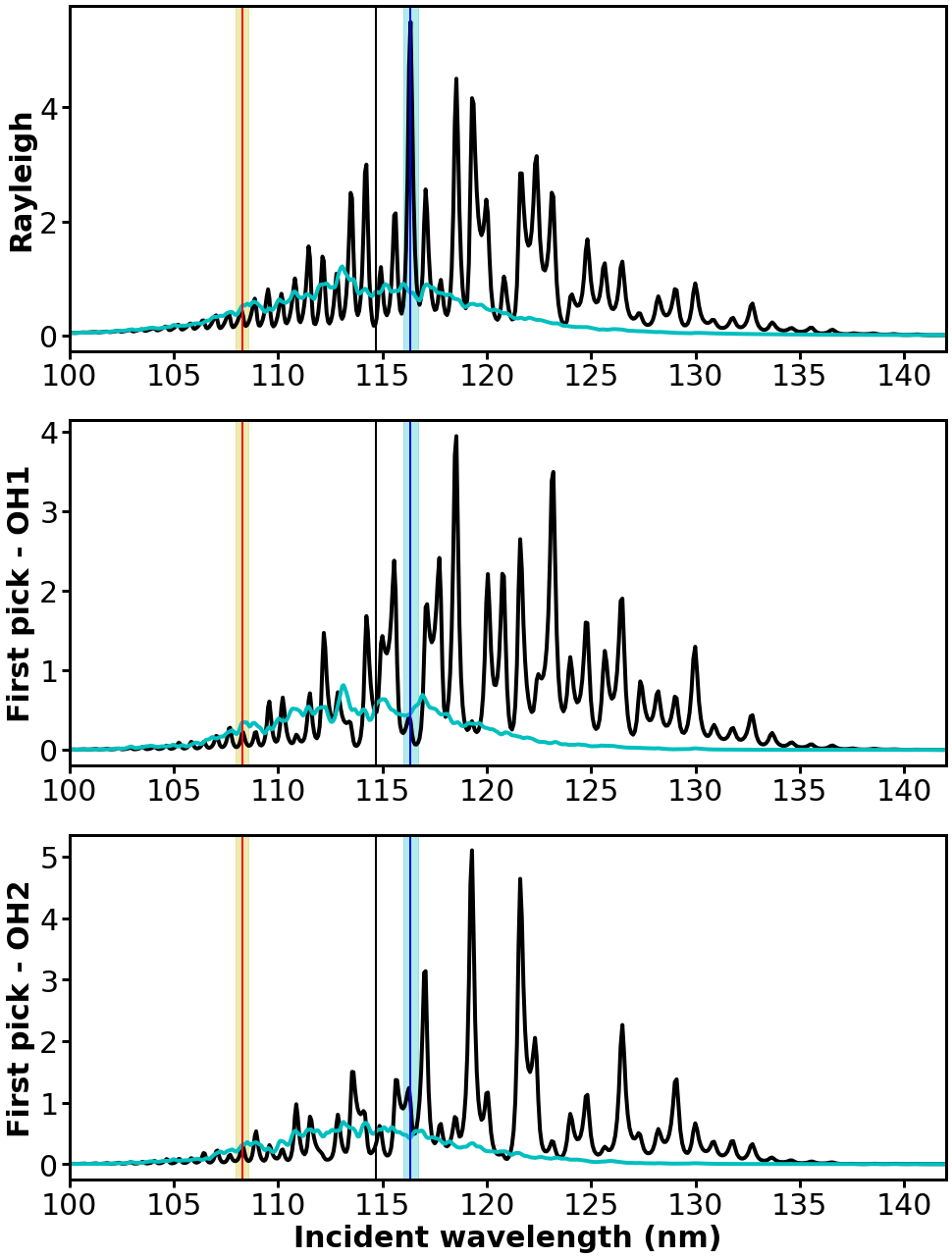}
\caption{\label{fig:37} Rayleigh intensity and Resonance Raman peak intensities of the IMDHO model 
with respect to the incident frequency for the LPI-MD approach (black curve), compared with the Placzek type polarizability (cyan curve), for different scattered frequencies: the Rayleigh peak (top panel), and the first peaks for both vibrational modes (middle and lower panels).}
\end{figure}

Finally, we investigate the difference between the LPI approach and a Placzek type approximation. To this end, we have calculated the RR spectrum approximating the frequency dependent polarizability by using solely the instantaneous excited state energy. The resulting Placzek type spectra are reported in Figures~\ref{fig:36} and~\ref{fig:37}. 
Figure~\ref{fig:36} shows that the overtones' intensities are too strong compared to the intensity of the fundamental modes. Similarly, the vibronic structure is completely lost in Palczek-type spectra, see Figure~\ref{fig:37}. LPI-MD thus appears as a clear improvement over this short-time approach.

We should  remind that for the IMDHO model 
the two linearization procedures that lead to the dynamics on the mean state for the Wigner transform of the polarizability on one hand, and the classical trajectory on the ground state, are exact. The only remaining approximation is the Wigner transform of product of operators written as product of Wigner transforms. Here we found that this approximation is  not too drastic as the rich RR spectrum of this simple model is remarkably well reproduced by the LPI-MD method. 
We will now turn to anharmonic systems.

\section{\label{sec:Heller}Heller's model}

\subsection{Description}

Here we will investigate how the linearization approach performs on a typical model used to study resonance Raman spectroscopy proposed by
by Heller, Sundberg and Tannor\cite{Heller82} some years ago and here
called simply \enquote{Heller's model}.
As for the IMDHO model, it is a two-dimensional model, now with an anharmonic excited state potential energy surface. More specifically, the ground and excited state potential energy surfaces are given by 
\begin{align}
  \begin{split}
    V_\mathrm{gs}(x,y)&=m\omega^2x^2/2+m\omega^2y^2/2\\
    V_\mathrm{es}(x,y)&=m\omega^2(x-x_0)^2/2+m\omega^2(y-y_0)^2/2\\
    &-K(x-x_0)^2(y-y_0)+V_0
  \end{split}
\end{align}

\begin{table}
\caption{\label{tab:01}Parameters used the Heller's model. }
\begin{tabular}{c | c c}
\textbf{Parameter}       & \textbf{Value} & \textbf{Units} \\
\hline
$x_0$       & 13.996                & pm \\
$y_0$       & 8.3978                & pm \\
$\omega$    & 1592.7                & cm$^{-1}$ \\
$V_{0}$     & 3.9493                & eV \\
$m$         & 6.7540                & g.mol$^{-1}$\\
$K$         & 1.2378$\cdot10^{-4}$  & eV.pm$^{-3}$
\end{tabular}
\end{table}

All parameters of the potential energy surfaces and used in the simulations are reported in Tables~\ref{tab:01} and~\ref{tab:03bis}, respectively.

In this model, the two vibrational modes along the two degrees of freedom are degenerate. The vibrational spectrum will then be composed of this fundamental frequency and their overtones. In particular we will be interested here, following Heller, in the intensities of the first overtone compared to the fundamental peak.


\begin{table}
\caption{\label{tab:03bis}Parameters used in the LPI-MD simulations with the Heller's model.}
\begin{tabular}{l | c}
Parameter & Value \\
\hline
Timestep & 0.1 fs \\
\#~Ground state trajectories & 2000 \\
\#~Mean state trajectories & 4500 \\
Length of the ground state trajectories & 1000 fs \\
Length of the mean state trajectories & 667 fs \\
Damping factor & 0.06 fs$^{-1}$\\
Scattered resolution & 0.03 rad fs$^{-1}$ \\
\end{tabular}
\end{table}

\subsection{Absorption and resonance Raman}

\begin{figure}[htbp]\centering
\includegraphics[width=\linewidth]{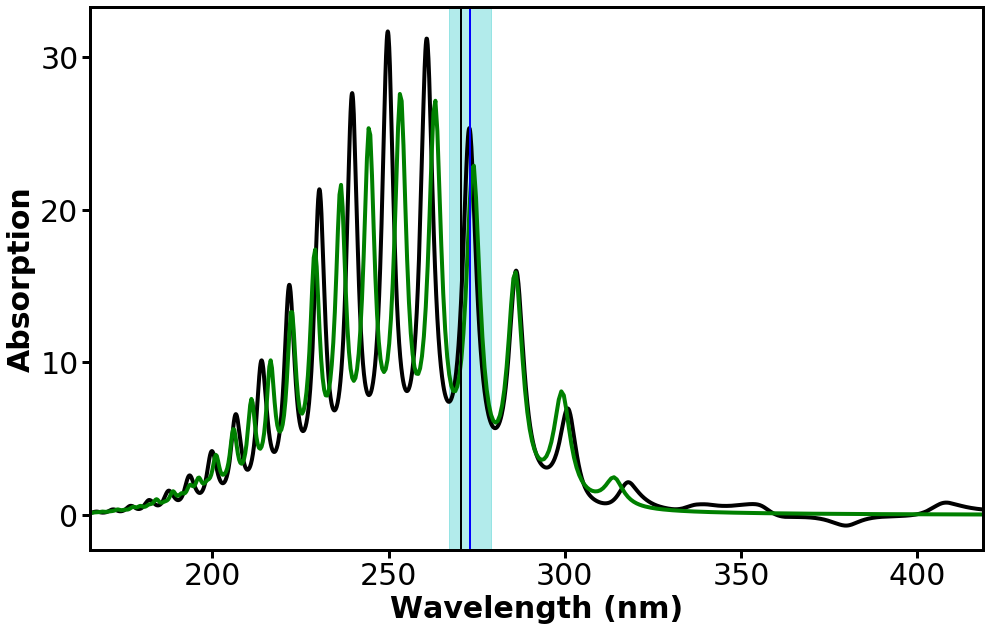}
\caption{\label{fig:19}Absorption spectra obtained from LPI-MD on Heller's model with $\Gamma = 0.02$ fs$^{-1}$ (black curve), compared with a \enquote{sum-over-states} approach (green curve). The near resonance is highlighted with a blue vertical line and a cyan integration area, while the Franck-Condon region with a black vertical line.}
\end{figure}

For this anharmonic model, the LPI-MD approach still reproduces well the overall shape and the vibronic structure of the absorption spectrum as can be seen in Figure~\ref{fig:19}. However, the spacing of vibronic structure appears slightly different with respect to the sum-over-states reference. 

\begin{figure}[htbp]\centering
\includegraphics[width=\linewidth]{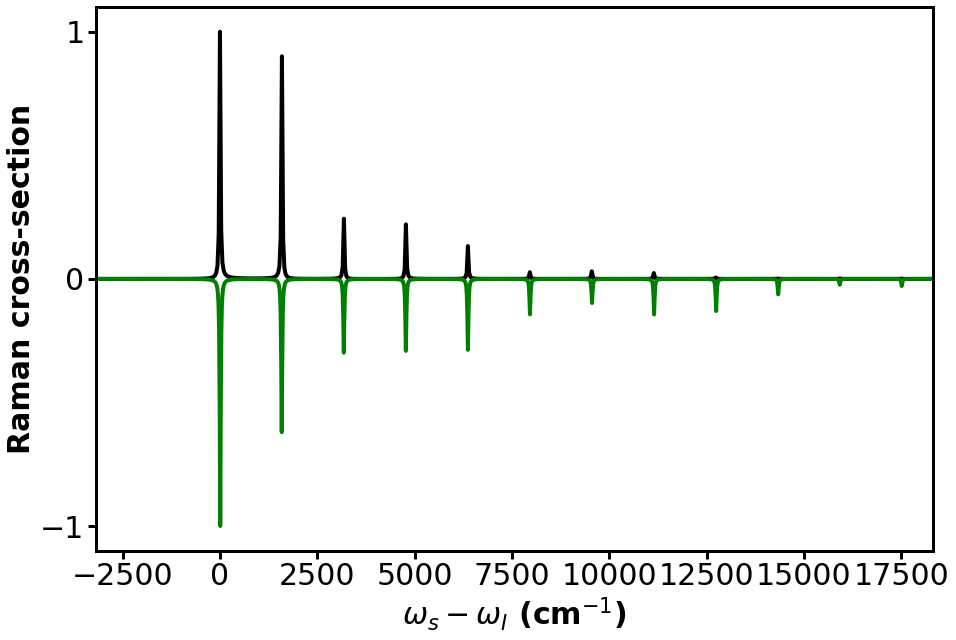}
\caption{\label{fig:23}Resonance Raman spectra for Heller's model with respect to the scattered frequency for the near-resonance incident frequency, compared with a sum-over-states approach (green curve). They are normalized with respect to the Rayleigh peak.}
\end{figure}

For the calculation of the resonance Raman spectrum, we have chosen an incident frequency corresponding to a vibronic peak which is in common between LPI-MD  and sum-over-states absorption spectra. Thus, the resonance Raman signals were obtained by integrating over the incident frequency around
275~nm (see Figure~\ref{fig:19}). 
The resulting signals are shown in Figure~\ref{fig:23}. Since the two oscillators have the same frequency, only one progression of overtones of this fundamental frequency is obtained. The peak intensities along this progression for the LPI-MD approach are very similar to the sum-over-state ones, showing that the proposed LPI-MD approach is able to correctly catch the non-linear character of the frequency dependent polarizablity.


\begin{figure}[htbp]\centering
\includegraphics[width=\linewidth]{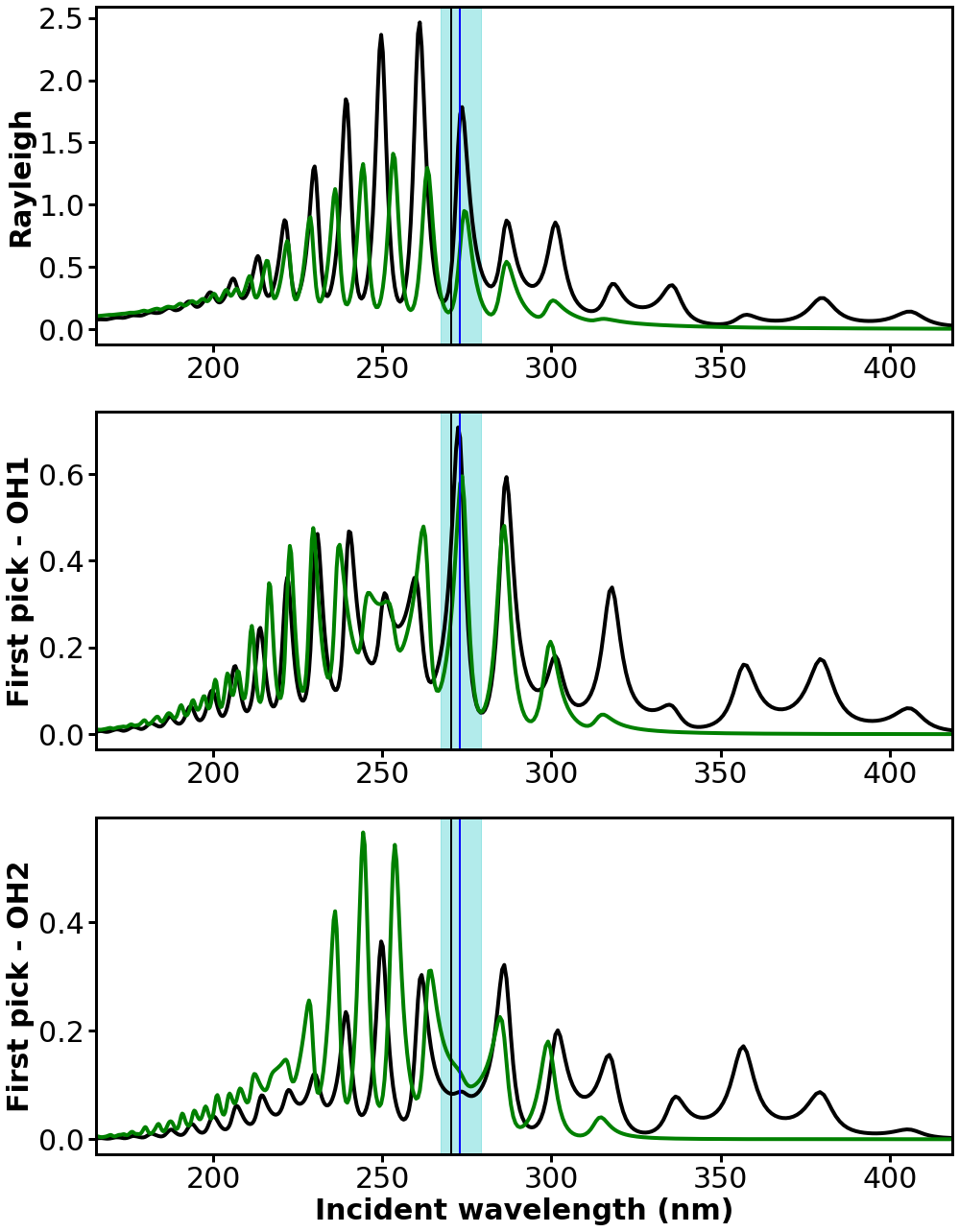}
\caption{\label{fig:21}Raman spectra for Heller model with respect to the incident frequency (black curve), compared with a sum-over-states approach (green curve), for different scattered frequencies: the Rayleigh pick (top), and first picks for both vibrational modes (middle and bottom), with highlights on relevant pulsation (blue) and Franck-Condon frequency (black).}
\end{figure}

We now focus on the intensity of the fundamental peak and first overtone as a function of the incident frequency, as shown in Figure~\ref{fig:21}. As for the absorption, the spacing of the vibronic bands is slightly larger than that given by the SoS reference. Nevertheless, the vibronic structure is well reproduced. 
It is interesting to see that at the frequency that corresponds to the highest intensity for the fundamental band, which had been chosen for plotting Figure~\ref{fig:23}, the overtone peak is quenched and this phenomenon observed in the SoS reference is well reproduced with the LPI approach. However, just like for the independent displaced modes, the LPI approach overestimates the scattered intensity at higher wavelengths of the incident light.


\section{\label{sec:HOCl}Hypochlorous acid}

\subsection{Description}

Finally, we investigate a model of a simple tri-atomic molecule, the hypochlorous
acid (HOCl), for which resonance Raman spectroscopy has proved to
be  a useful tool to study also 
the excited state potential energy surface\cite{molina78,hickman93}. In fact, the system
has a dissociating channel in the $\tilde{X}^1A'$ excited state, 
corresponding to the formation of OH ($^2\Pi$) and Cl ($^2P$)\cite{molina78,nambu89,hickman93}.
This system was previously studied theoretically by using time-dependent
approaches\cite{nanbu92,Offer96}, based on simple analytical Hamiltonian. In our study, we have used the model proposed by Nambu and Iwata\cite{nanbu92} modified on the ground state bending vibration to
better reproduce the known vibrational frequency\cite{molina78}.


The ground state potential is thus described by a Morse potential for the stretching ($R$) and a harmonic oscillator for the bending ($\theta$):

\begin{equation}
\label{eq:HOClV}
    V^g(R,\theta) = D_e\times(1-e^{-a(R-R_e)})^2 + \frac{k_{\theta}}{2}(\theta-\theta_e)^2
\end{equation}

The system has a dissociating channel in the excited state, which is
described by a generalized anti-Morse potential\cite{Sato1955a,Sato1955b}
to describe the diffuse
state of the O--Cl stretching:

\begin{subequations}
\begin{equation}
V^{ex}(R,\theta)=P_1+(P_4-P_1)[1+\exp\{-P_3(R-P_2)\}]^2
\end{equation}
\begin{equation}
P_{i \neq 2}=\sum_{j=0}^3C_{ij}\theta^{2j}
\end{equation}
\end{subequations}

with $P_2$ is fixed at $120\,\text{pm}$ and 
where $R$ represents the O--Cl distance (we disregard the small offset of
the OH center of mass from O atom as in the original work of Nanbu and Iwata\cite{nanbu92}) and $\theta$ the HOCl angle.
The O--H distance is kept fixed in the model to the equilibrium 
experimental value of 0.9643~\mbox{\AA} and thus it is not considered in the dynamics\cite{nanbu92}.
All the parameters used to describe the ground and excited potential 
energy surfaces are reported in Table~\ref{tab:HOCl_pot} and in the
original work of Nanbu and Iwata\cite{nanbu92}. 

\begin{table}[]
    \caption{Parameters used to define the ground state potential 
    of HOCl (see Eq.~\eqref{eq:HOClV}) and kinetic energy (see Eq.~\eqref{eq:HOClK}).}
    \begin{tabular}{l | r}
Parameter & Value \\
\hline
$R_e$             & 168.91~pm \\
$\theta_e$        & $\ang{102.45}$ \\
$\mu$             & 11.49~Da \\
$I$               & $8524\,\text{Da.pm}^2$ \\
$D_e$             & 1.92~eV \\
$a$               & 0.02444~pm$^{-1}$ \\
$k_{\theta}$      & $4.977\,\text{eV.rad}^{-2}$ \\
$\omega_R$        & 738~cm$^{-1}$ \\
$\omega_{\theta}$ & 1272~cm$^{-1}$ \\
    \end{tabular}
    \label{tab:HOCl_pot}
\end{table}



Since the present aim is not to obtain the exact HOCl spectrum, but to test our LPI-MD approach on a realistic mode, we made two approximations to simplify the description: (i) we ignored the rotation of the molecule and considered only $J=0$ states; (ii) the Herzberg-Teller and polarization effects have being neglected, assuming equal and constant transition moments along the two modes.

The Hamiltonian used in the simulations is thus:
\begin{equation}\label{eq:HOClK}
    H_{HOCl} = \frac{P^2}{2\mu}+\frac{J_\theta^2}{2I}+V(R,\theta)
\end{equation}
where $I$ is the moment of inertia associated to the angle $\theta$.
The parameters used in LPI-MD simulations are summarized in Table~\ref{tab:04}.

\begin{table}
\caption{\label{tab:04}Constants used for the propagation of HOCl LPI-MD simulations. }
\begin{tabular}{l | c}
Parameter & Value \\
\hline
Timestep    & 0.1~fs \\
\#~Ground state trajectories & 2000 \\
\#~Mean state trajectories & 10~000 \\
Length of the ground state trajectories & 3000~fs \\
Length of the mean state trajectories & 150~fs \\
Damping factors $\Gamma$& 0.09 -- 0.5~fs$^{-1}$ \\
\hline
Scattered resolution & 0.04~fs$^{-1}$ \\
\end{tabular}
\end{table}

\subsection{Absorption and damping factor}

\begin{figure}[htbp]\centering
\includegraphics[width=\linewidth]{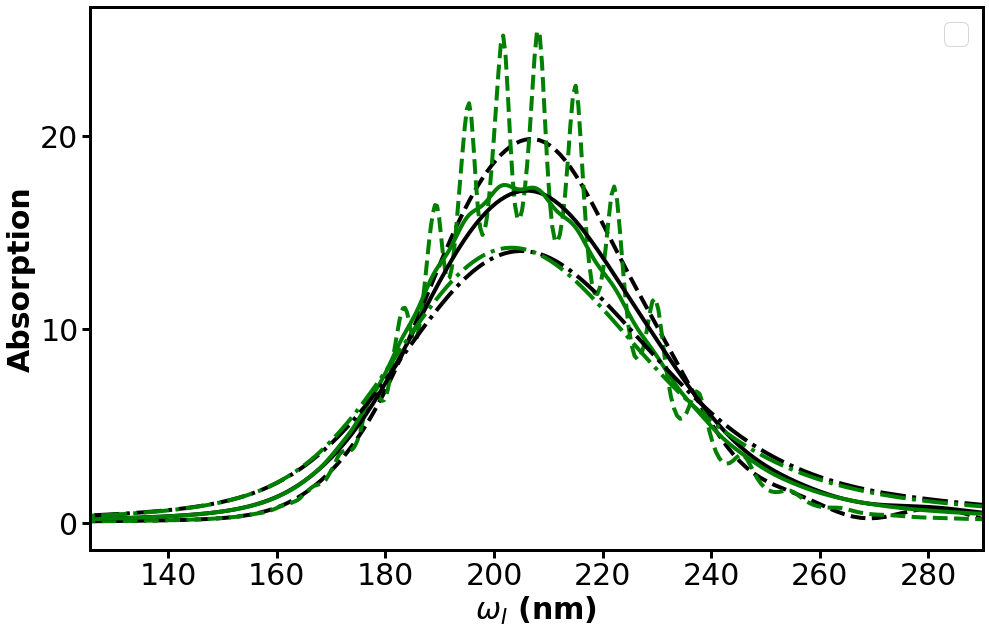}
\caption{\label{fig:26}Absorption spectra for HOCl model with $\Gamma = 0.09$ fs$^{-1}$ (dashed curves), $\Gamma = 0.25$ fs$^{-1}$ (full curves), $\Gamma = 0.5$ fs$^{-1}$ (half-dashed curves) -- black curves, compared with a \enquote{sum-over-states} approach (green curves).}
\end{figure}

Figure~\ref{fig:26} shows the calculated absorption spectrum for different values of the phenomenological damping factor in the 0.09--0.5~fs$^{-1}$ range. In the sum-over-states calculations, we have used a confinement potential to obtain excited-states vibrational eigenstates, due to the
dissociative character of the potential energy surface. This results in an artificial quantization of these vibrational eigenstates while a continuum of states should be present. The corresponding absorption spectrum then shows a vibronic structure at low $\Gamma$ that disappears when $\Gamma$ is increased, mimicking a broadening of the energy eigenvalue similar to the expected continuum spectrum. 

\begin{figure}[htbp]\centering
\includegraphics[width=\linewidth]{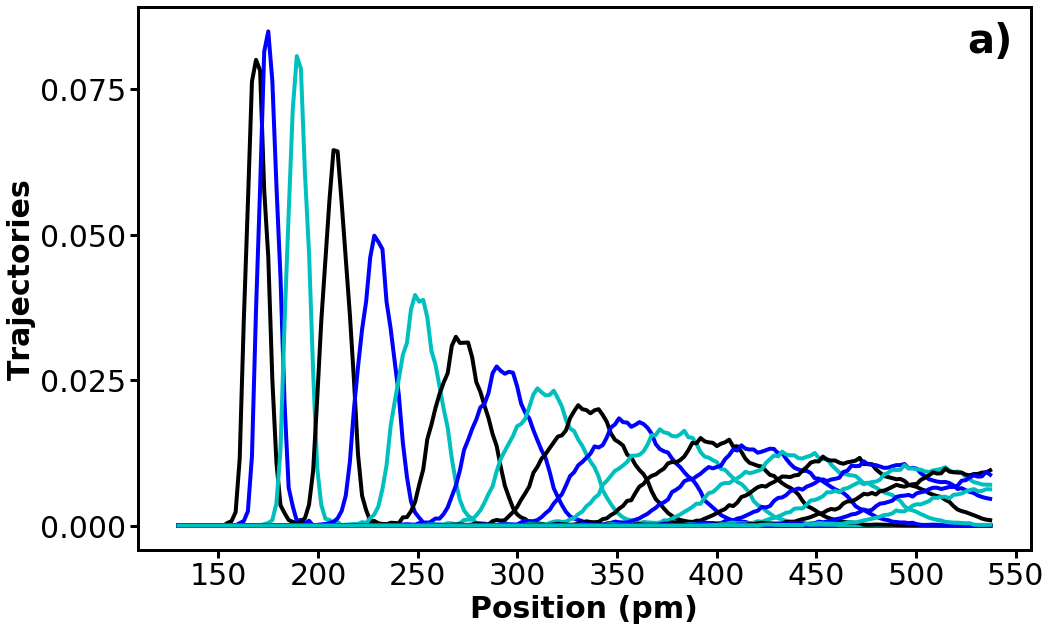}
\includegraphics[width=\linewidth]{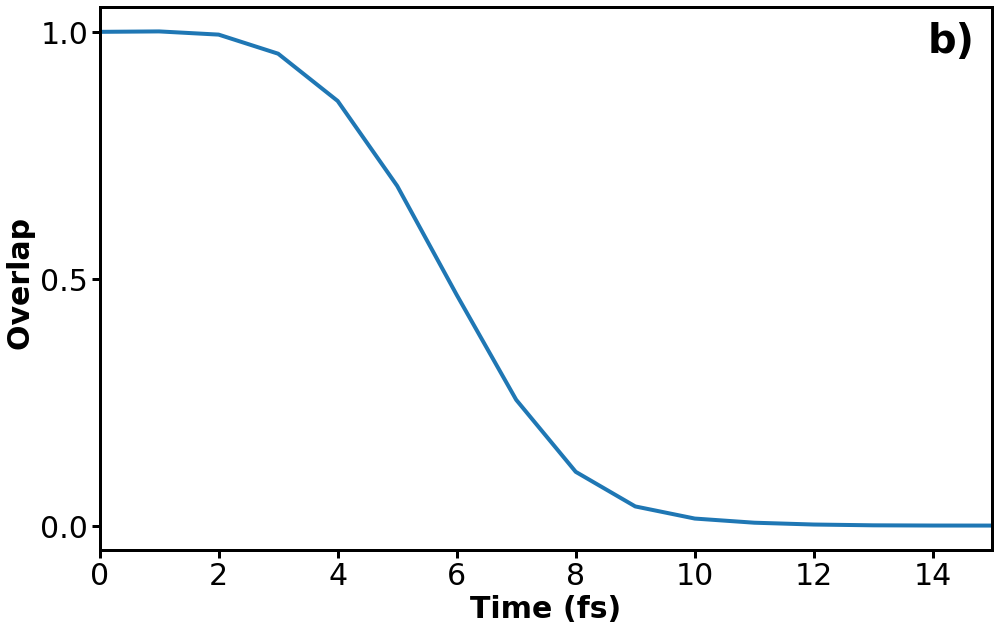}
\caption{\label{fig:24}Behavior of R coordinate in HOCl trajectories over the mean PES. Panel a): R distributions at different time frames. 
Colors are only meant to help reading and curves are each $5$ fs distant.
Panel b): time evolution of the histograms' overlaps.}
\end{figure}

On the other hand, the LPI-MD simulations provide an absorption spectrum
nearly independent of the damping factor (see black curves in the same
Figure~\ref{fig:26}). 
This is due to the dissociative trajectories on the mean potential energy surface. Figure~\ref{fig:24}.a shows the histograms of R coordinate at different 
times along the trajectory on the mean PES, averaged over all initial positions and momenta. 
It clearly appears that these trajectories are dissociating 
and the time-scale can be inferred from the overlap of the distribution at time $t$ and the initial distribution. This time-dependent overlap is displayed on Figure~\ref{fig:24}.b and shows a typical timescale for dissociation on the mean surface of around 6~fs.  

Coming back to the absorption spectrum, we find that a good agreement between the sum-over-states approach and the LPI-MD method is found for $\Gamma=0.25$~fs$^{-1}$ which would correspond to a characteristic time of 4~fs, which is on the same order of magnitude of the lifetime observed 
from the dynamics on the mean PES. Lower values of $\Gamma$ lead to an artificial vibronic structure in the sum-over-states approach, while large values of $\Gamma$ tend naturally to wash away any structure in the spectrum.
 
\subsection{Resonance Raman}

\begin{figure}[htbp]\centering
\includegraphics[width=\linewidth]{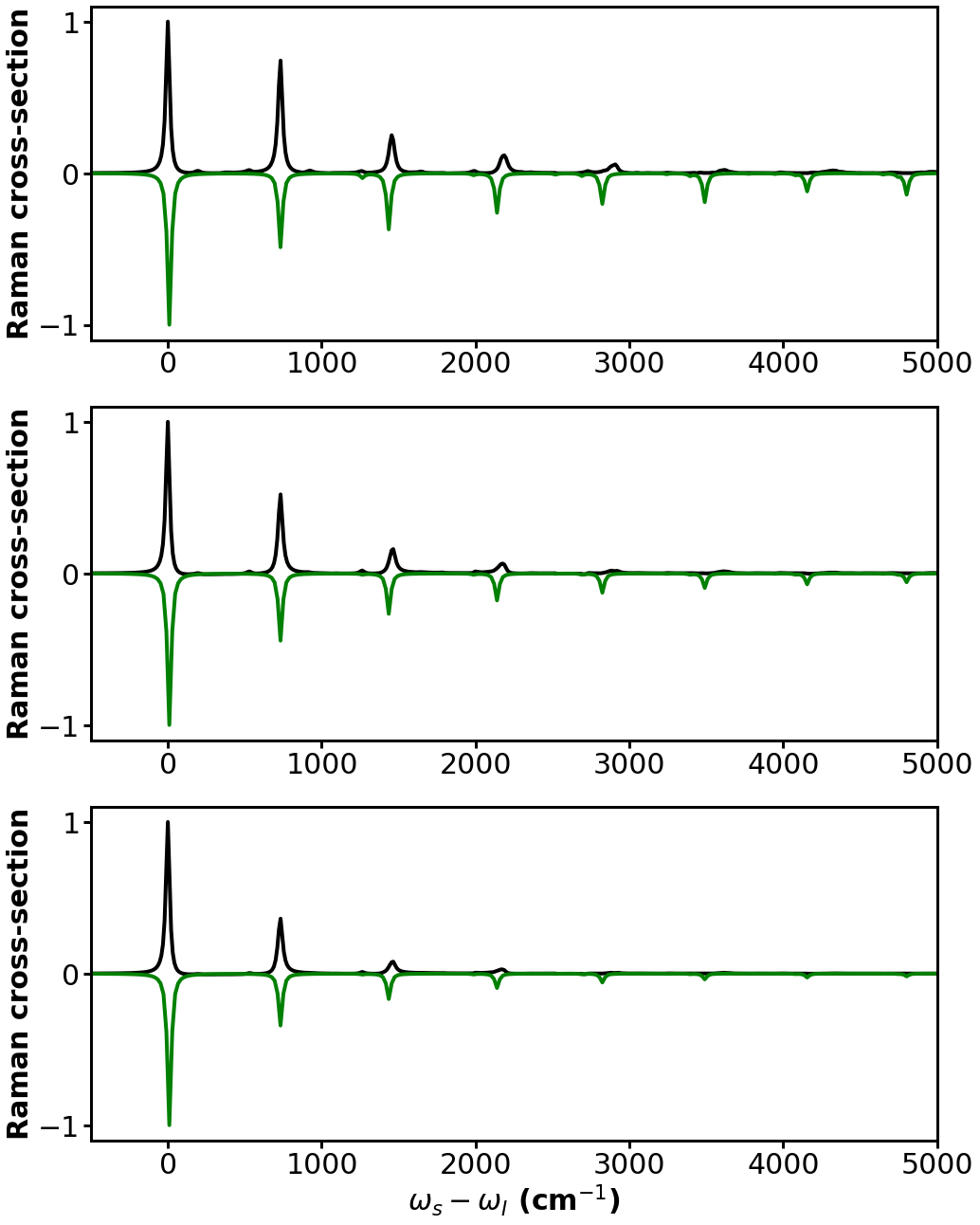}
\caption{\label{fig:29}Raman spectra for HOCl model with respect to the scattered frequency for the incident frequencies highlighted in Figure \ref{fig:27}  with $\Gamma = 0.09$ fs$^{-1}$ (top), $\Gamma = 0.25$ fs$^{-1}$ (middle), $\Gamma = 0.5$ fs$^{-1}$ (bottom) -- black curves, compared with a \enquote{sum-over-states} approach (green curves), scaled to have the same intensity for the Rayleigh pick.}
\end{figure}

In Figure~\ref{fig:29} we show the RR spectrum using an incident
frequency at resonant conditions at 200~nm 
obtained from LPI-MD simulations and compared with SoS results.
The Raman cross-section, normalized with respect to the Rayleigh
peak, is shown for three values of the damping factor, $\Gamma$.
The overall agreement is very good in all three cases, although the overtones' intensities appear to decrease more quickly 
in the LPI-MD simulations than in the SoS ones, similar to what is observed in the Heller model. 

\begin{figure}[htbp]\centering
\includegraphics[width=\linewidth]{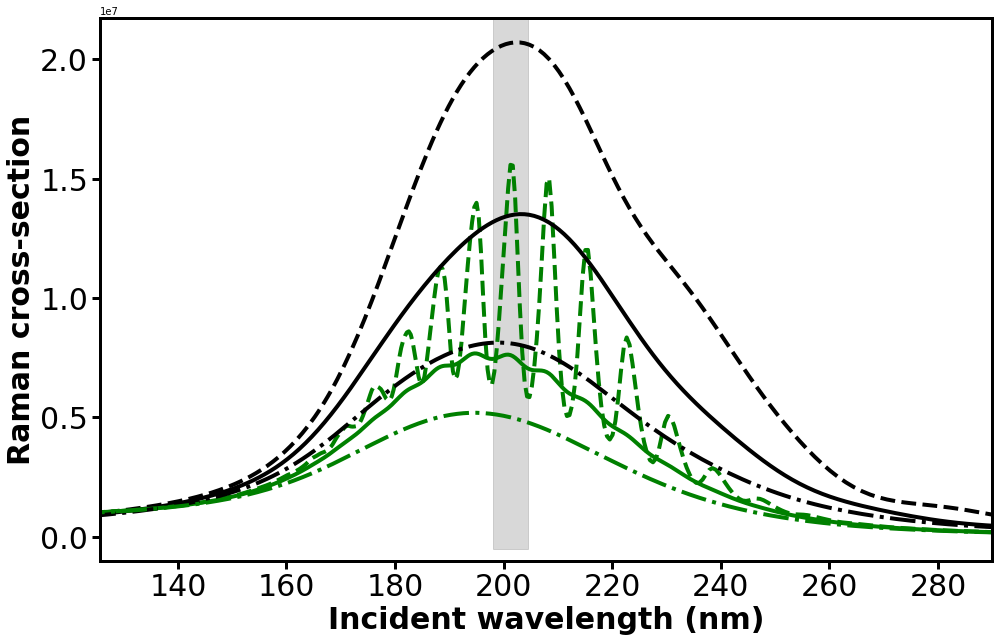}
\caption{\label{fig:27}Raman cross-section as a function of the incident wavelength obtained from the HOCl model for the Rayleigh peak for  $\Gamma = 0.09$ fs$^{-1}$ (dashed lines), $0.25$ fs$^{-1}$ (full lines), $0.5$ fs$^{-1}$ (dot-dashed lines). LPI-MD results are in black, while SoS ones are in green. The wavelength near-resonance region used for scattered resonance Raman spectra is marked in grey.}
\end{figure}

Finally, we investigated the intensity of the first resonance Raman peak as a function of the incident frequency. The result is shown in Figure~\ref{fig:27} for different  damping factors. Similar to the absorption, $\Gamma=0.25$~fs$^{-1}$ leads to a good agreement between LPI-MD and sum-over-states results, with only a constant scaled intensity.  The agreement is still good for the larger value of $\Gamma$ while 
for the lower value we obtain a too strong vibronic structure in the
SoS calculations. As discussed previously, this is due to the unphysical bounding potential added in the SoS calculations. 

\section{\label{sec:conc}Conclusion}

In this work, we propose a new method to obtain resonance Raman 
spectra based on path-integral linearization and amenable to molecular dynamics simulations.
The full derivation leads to a relatively simple algorithm which in 
principle can be applied to atomistic simulations. 
The central quantity 
is the time evolution of the ground-to-excited state energy difference during dynamics on the mean potential energy surface. Such dynamics is typical of linearization techniques\cite{Bonella2005,Bonella2005b}. This is used to compute a time-dependent polarizability whose auto-correlation function leads to the RR signal. This extends upon previous work based on Placzek type polarizability and includes vibronic effects and enhancement of overtones and combination bands intensities in the RR spectrum.
In the present 
work we have tested the method on three models: harmonic and 
anharmonic potentials plus a model of HOCl with a dissociating excited
state. The comparison with sum-over-states results show that the method
globally catch the most important features. 
For independent mode displaced harmonic oscillators, the model is by construction nearly exact and this is shown by comparison to SoS results. 

With the advent of efficient methods to compute excited states energies and gradients using density functional theory\cite{Mattiat2021}  it is possible to envision the simulation of RR spectra beyond the short-time Placzek approximation for realistic systems. Also, to this aim, one could make use of recent developments, based on quantum thermal bath\cite{Dammak2009,Barrat2011,Brieuc2016,Mangaud2019,Mauger2021}, to generate dynamics that preserve the Wigner distribution thus allowing to perform one single long simulation instead of sampling initial conditions while retaining the quantum description of vibrational modes\cite{Bonella2021}.

Concluding, this work gives the basis to proceed further and be able to get accurate 
resonance Raman spectra (with fundamentals, overtones and combination bands) for 
real systems, since the present approach can be extended to molecular dynamics simulations in 
a relatively simple way. Our research is going in this direction.

\begin{acknowledgments}
\end{acknowledgments}
We thank Dr. Alberto Mezzetti for useful discussions in particular in raising
our interest for resonance Raman in biophysical chemistry. We are also grateful to Dr. Sara Bonella for her invaluable insight on the linearization approach for the computation of quantum time-correlation functions.

\appendix

\section{Linearization of the polarizability operator}\label{sec:LPI}

Here we show how we can apply the linearization path integral (LPI) procedure to the Wigner transform of polarizability operator of Eq.~\ref{eq:pol2app}.
In particular we can re-write the second term in Eq.~\ref{eq:pol2app} as:

\begin{align}
\begin{split}\label{eq:propapp1}
    \braket{x-\frac{\Delta x}{2}}{\widehat{M_s}e^{-\frac{j}{\hbar}t\op{H_b}}\widehat{M_I}e^{\frac{j}{\hbar}t\op{H_a}}}{x+\frac{\Delta x}{2}}&=\left<x-\frac{\Delta x}{2}\middle|\widehat{M_s}\middle|x^+_{P+1}\middle>\middle<x^+_{P+1}\middle|e^{-\frac{j}{\hbar}t\op{H_b}}\middle|x^+_{0}\right>\\
    &\times\left<x^+_{0}\middle|\widehat{M_I}\middle|x^-_{0}\middle>\middle<x^-_{0}\middle|e^{\frac{j}{\hbar}t\op{H_a}}\middle|x^-_{P+1}\right>
\end{split}
\end{align}

where we can identify the term $x^-_{P+1}$ with $x+\frac{\Delta x}{2}$.
We can now re-write using the path-integral formalism where
$e^{\frac{j}{\hbar}t \widehat{H}} = \left( e^{\frac{j}{\hbar}\epsilon \widehat{H}} \right)$, with $\epsilon = \frac{t}{P+1}$, such that the two terms in
Eq.~\ref{eq:propapp1} which contain the propagators on surfaces $a$ and $b$ are:

\begin{align}\label{eq:perappendix}
\begin{split}
    \braket{x_0^-}{e^{\frac{j}{\hbar}t \op{H_a}}}{x^-_{P+1}}\braket{x^+_{P+1}}{e^{- \frac{j}{\hbar}t \op{H_b}}}{x_0^+}&=\intgm{}{x_1^+}{x_P^+}{}\intgm{}{x_1^-}{x_P^-}{}\\
    &\times\braket{x^+_{P+1}}{e^{-\frac{j}{\hbar}\varepsilon \op{H_b}}}{x_P^+}\dots \braket{x_1^+}{e^{-\frac{j}{\hbar}\varepsilon \op{H_b}}}{x_0^+}\\
    &\times\braket{x^-_{0}}{e^{\frac{j}{\hbar}\varepsilon \op{H_a}}}{x_1^-}\dots \braket{x_P^-}{e^{\frac{j}{\hbar}\varepsilon \op{H_a}}}{x_{P+1}^-}
\end{split}
\end{align}

We can now apply the Trotter's theorem, such that

\begin{align}
\begin{split}
    &\left<x^+_{k+1}\middle|e^{- \frac{j}{\hbar}\varepsilon \widehat{{H}_b}}\middle|x_k^+\right> = \left<x^+_{k+1}\middle|e^{- \frac{j}{\hbar}\frac{\varepsilon}{2} \widehat{V_b}}e^{- \frac{j}{\hbar}\varepsilon \frac{\widehat{p}^2}{2m}}e^{- \frac{j}{\hbar}\frac{\varepsilon}{2} \widehat{V_b}}\middle|x_k^+\right>\\
    &=e^{-\frac{j}{\hbar}\frac{\varepsilon}{2} V_b(x^+_{k+1})}
    \left<x^+_{k+1}\middle|
    e^{- \frac{j}{\hbar}\varepsilon \frac{\widehat{p}^2}{2m}}\middle|x_k^+\right>e^{-\frac{j}{\hbar}\frac{\varepsilon}{2} V_b(x_k^+)}\\
    &= e^{-\frac{j}{\hbar}\frac{\varepsilon}{2} V_b(x^+_{k+1})}\int_{-\infty}^{+\infty}dp^+_{k+1}\left<x^+_{k+1}\middle|e^{- \frac{j}{\hbar}\varepsilon \frac{\widehat{p_2}^2}{2m}}\middle|p^+_{k+1}\right>\left<p^+_{k+1}\middle|x_k^+\right>e^{-\frac{j}{\hbar}\frac{\varepsilon}{2} V_b(x^+_k)}\\
    &=\frac{1}{2\pi\hbar}\int_{-\infty}^{+\infty}dp^+_{k+1}e^{\frac{j}{\hbar}p_{k+1}^+\left(x_{k+1}^+-x^+_k\right)}e^{-\frac{j}{\hbar}\frac{\left(p_{k+1}^+\right)^2}{2m}}e^{-\frac{j}{\hbar}\frac{\varepsilon}{2} V_b(x^+_{k+1})}e^{-\frac{j}{\hbar}\frac{\varepsilon}{2} V_b(x^+_k)}
\end{split}
\end{align}

and similarly on surface $a$
\begin{align}
\begin{split}
    &\left<x^-_{k}\middle|e^{\frac{j}{\hbar}\varepsilon \widehat{{H}_a}}\middle|x_{k+1}^-\right>=\left(\left<x^-_{k+1}\middle|e^{- \frac{j}{\hbar}\varepsilon \widehat{{H}_a}}\middle|x_k^-\right>\right)^*\\
    &=\frac{1}{2\pi\hbar}\int_{-\infty}^{+\infty}dp_{k+1}^-e^{-\frac{j}{\hbar}p_{k+1}^-\left(x_{k+1}^--x^-_k\right)}e^{\frac{j}{\hbar}\frac{\left(p_{k+1}^-\right)^2}{2m}}
    e^{\frac{j}{\hbar}\frac{\varepsilon}{2} V_a(x^-_{k+1})}e^{\frac{j}{\hbar}\frac{\varepsilon}{2} V_a(x^-_k)}
\end{split}
\end{align}

Note that for $n$ dimensional coordinates the prefactor is then to be put to the power $n$. We thus get:

\begin{align}
\label{eq:pathintB}
\begin{split}
    \left<x_0^-\middle|e^{\frac{j}{\hbar}t \widehat{{H}_a}}\middle|x^-_{P+1}\right>&\left<x^+_{P+1}\middle|e^{- \frac{j}{\hbar}t \widehat{{H}_b}}\middle|x_0^+\right>=\frac{1}{(2\pi\hbar)^{2(P+1)}}\int_{-\infty}^{+\infty}\!\!\!\!\!\!\!\!\dots \int_{-\infty}^{+\infty}\!\!\!\!dx_1^+\!\!\dots dx_{P}^+ \int_{-\infty}^{+\infty}\!\!\!\!\!\!\!\!\dots \int_{-\infty}^{+\infty}\!\!\!\!dx_1^-\!\!\dots dx_{P}^-\\
    &\times\int_{-\infty}^{+\infty}\!\!\!\!\!\!\!\!\dots \int_{-\infty}^{+\infty}\!\!\!\!dp_1^+\!\!\dots dp_{P}^+ \int_{-\infty}^{+\infty}\!\!\!\!\!\!\!\!\dots \int_{-\infty}^{+\infty}\!\!\!\!dp_1^-\!\!\dots dp_{P}^-\\
    &\exp\left[\frac{j}{\hbar}\sum_{k=0}^{P}\left(p^+_{k+1}(x^+_{k+1}-x^+_k)-p^-_{k+1}(x^-_{k+1}-x^-_k)\right)\right]\\
    &\times\exp\left[-\frac{j}{\hbar}\varepsilon\sum_{k=0}^{P}\left(\frac{(p_{k+1}^+)^2}{2m}-\frac{(p_{k+1}^-)^2}{2m}\right)\right]
    \exp\left[-\frac{j}{\hbar}\frac{\varepsilon}{2}\sum_{k=0}^{P}\left(V_a(x_k^+)-V_b(x_k^-)\right)\right]\\
    &\times\exp\left[-\frac{j}{\hbar}\frac{\varepsilon}{2}\sum_{k=1}^{P}\left(V_a(x_k^+)-V_b(x_k^-)\right)\right]
\end{split}
\end{align}

In view of performing linearization with respect to the difference between backward and forward paths, we first make a change of variables:
\begin{equation}
    \overline{y}=\frac{y^++y^-}{2}\quad;\quad\Delta y=y^+-y^-\quad\left(\text{det}(J_F)=-1\right),
\end{equation}
where $y\equiv x\text{ or }p$.
We then have the following relations
\begin{subequations}
\begin{equation}
    (p^+_{k+1})^2-(p^-_{k+1})^2=2\overline{p_{k+1}}\Delta p_{k+1}
\end{equation}
\begin{align}
  \begin{split}
    \sum_{k=0}^P&\left(p^+_{k+1}(x^+_{k+1}-x^+_k)-p^-_{k+1}(x^-_{k+1}-x^-_k)\right)\\
    &=\sum_{k=0}^P\Delta p_{k+1}\left(\overline{x_{k+1}}-\overline{x_k}\right)-\sum_{k=1}^P\Delta x_k\left(\overline{p_{k+1}}-\overline{p_k}\right)\\
    &-\overline{p_1}\Delta x_0+\overline{p_{P+1}}\Delta x_{P+1}.
  \end{split}
\end{align}
\end{subequations}

Thus, we can  re-write Eq.~\eqref{eq:pathintB} as:
\begin{widetext}
\begin{align}
\begin{split}
    \left<x_0^-\middle|e^{\frac{j}{\hbar}t \widehat{{H}_a}}\middle|x^-_{P+1}\right>&\left<x^+_{P+1}\middle|e^{- \frac{j}{\hbar}t \widehat{{H}_b}}\middle|x_0^+\right>=\frac{1}{(2\pi\hbar)^{2(P+1)}}\\
    &\int_{-\infty}^{+\infty}\!\!\!\!\!\!\!\!\dots \int_{-\infty}^{+\infty}\!\!\!\!d\overline{x_1}\!\!\dots d\overline{x_P} \int_{-\infty}^{+\infty}\!\!\!\!\!\!\!\!\dots \int_{-\infty}^{+\infty}\!\!\!\!d\overline{p_1}\!\!\dots d\overline{p_{P+1}}\exp\left[\frac{j}{\hbar}\left(\overline{p_{P+1}}\Delta x_{P+1}-\overline{p_1}\Delta x_0\right)\right]\\
    &\times \int_{-\infty}^{+\infty}\!\!\!\!\!\!\!\!\dots \int_{-\infty}^{+\infty}\!\!\!\!d\Delta x_1\!\!\dots d\Delta x_P\\
    &\exp\left[-\frac{j}{\hbar}\sum_{k=1}^P\Delta x_k\left(\overline{p_{k+1}}-\overline{p_k}\right)\right]\exp\left[-\frac{j}{\hbar}\varepsilon\sum_{k=0}^PV_b\left(\overline{x_k}+\frac{\Delta x_k}{2}\right)-V_a\left(\overline{x_k}-\frac{\Delta x_k}{2}\right)\right]\\
    &\times \int_{-\infty}^{+\infty}\!\!\!\!\!\!\!\!\dots \int_{-\infty}^{+\infty}\!\!\!\!d\Delta p_1\!\!\dots d\Delta p_{P+1}\exp\left[-\frac{j}{\hbar}\varepsilon\sum_{k=0}^P \frac{\overline{p_{k+1}}\Delta p_{k+1}}{m}\right]\exp\left[\frac{j}{\hbar}\sum_{k=0}^P\Delta p_{k+1}\left(\overline{x_{k+1}}-\overline{x_k}\right)\right]
\end{split}
\end{align}
\end{widetext}
Integrating over the variables $\Delta p_k$, we obtain Dirac distributions  such that the propagation of the mean positions is driven by the momentum in the mean path:
\begin{align}
\begin{split}
    \int_{-\infty}^{+\infty}\!\!\!\!\!\!\!\!\dots \int_{-\infty}^{+\infty}\!\!\!\!&d\Delta p_1\!\!\dots d\Delta p_{P+1}\exp\left[-\frac{j}{\hbar}\varepsilon\sum_{k=0}^{P} \frac{\overline{p_{k+1}}\Delta p_{k+1}}{m}\right]\\
    &\times\exp\left[\frac{j}{\hbar}\sum_{k=0}^{P}\Delta p_{k+1}\left(\overline{x_{k+1}}-\overline{x_k}\right)\right]\\
    &=(2\pi\hbar)^{P}\prod_{k=0}^{P}\delta\left(\varepsilon \frac{\overline{p_{k+1}}}{m}-\overline{x_{k+1}}+\overline{x_k}\right)
\end{split}
\end{align}

To obtain the evolution of the momentum in the mean path from a force, we now apply linearization by assuming that $\delta x_k$ is small:
\begin{subequations}
\begin{align}\label{eq:linearization}
  \begin{split}
    V_b\left(\overline{x_k}+\frac{\Delta x_k}{2}\right)&-V_a\left(\overline{x_k}-\frac{\Delta x_k}{2}\right)\\
    &\approx V_b\left(\overline{x_k}\right)-V_a\left(\overline{x_k}\right)+\Delta x_k\nabla V_m(\overline{x_k})
  \end{split}
\end{align}
where we intruduce the mean potential energy surface $V_m$:
\begin{equation}
    V_m=\frac{V_a+V_a}{2}
\end{equation}
\end{subequations}

By integrating over $\Delta x_k$ in this linearized expression, we again obtain a series of Dirac distributions:
\begin{align}
\begin{split}
    \int_{-\infty}^{+\infty}\!\!\!\!\!\!\!\!&\dots \int_{-\infty}^{+\infty}\!\!\!\!d\Delta x_1\!\!\dots d\Delta x_{P+1}\exp\left[-\frac{j}{\hbar}\sum_{k=1}^P\Delta x_k\left(\overline{p_{k+1}}-\overline{p_k}\right)\right]\\
    &\times\exp\left[-\frac{j}{\hbar}\varepsilon\sum_{k=0}^PV_b\left(\overline{x_k}+\frac{\Delta x_k}{2}\right)-V_a\left(\overline{x_k}-\frac{\Delta x_k}{2}\right)\right]\\
    &=(2\pi\hbar)^P\prod_{k=1}^P\delta\left(\overline{p_{k+1}}-\overline{p_k}+\varepsilon\nabla V_m(\overline{x_k})\right)\\
    &\times\exp\left[-\frac{j}{\hbar}\varepsilon\sum_{k=0}^PV_b\left(\overline{x_k}\right)-V_a\left(\overline{x_k}\right)\right]
\end{split}
\end{align}


Now Eq.~\eqref{eq:pathintB} can be written in a compact form

\begin{align}
  \begin{split}
    \braket{x_{P+1}^+}{e^{-\frac{j}{\hbar}t\op{H_b}}}{x_0^+}\braket{x_0^-}{e^{\frac{j}{\hbar}t\op{H_a}}}{x_{P+1}^-}&=\frac{1}{2\pi\hbar}\int_{-\infty}^{+\infty}\left\{\begin{matrix}d\overline{p_1}\delta\gpar{\varepsilon\frac{\overline{p_{P+1}}}{m}-\overline{x_{P+1}}+\overline{x_P}}\\d\overline{p_{P+1}}\delta\gpar{\varepsilon\frac{\overline{p_{1}}}{m}-\overline{x_{1}}+\overline{x_0}}\end{matrix}\right.\\
    &\times\exp\gcro{\frac{j}{\hbar}\gpar{\overline{p_{P+1}}\Delta x_{P+1}-\overline{p_1}\Delta x_0}}\\
    &\times\exp\gcro{-\frac{j}{2\hbar}\varepsilon\sum_{k=0}^P
    \begin{pmatrix*}[l]
    V_b\gpar{\overline{x_k}}-V_a\gpar{\overline{x_k}}\\
    +V_b\gpar{\overline{x_{k+1}}}-V_a\gpar{\overline{x_{k+1}}}\\
    \end{pmatrix*}}
  \end{split}
\end{align}

where the trajectory on a mean surface $V_m=\frac{V_a+V_b}{2}$ 
can start either from $\left(\overline{x_0},\overline{p_1}\right)$ or from $\left(\overline{x_{P+1}},\overline{p_{P+1}}\right)$ but backward in time.

\section{Generalization to N excited states}\label{sec:generalb}
Here we report the equations to calculate Resonance Raman and absorption
spectra for a general system composed by $N$ excited states that in the
manuscript are reported only for one excited state $b$.

Wigner transform of the polarizability operator and its adjoint:

\begin{equation}\label{eq:wigpola_bis}
    \gpar{\op{\mathcal{P}}(\omega_I)}_W[x,p]=\frac{j}{\hbar} \sum_b\intg{e^{-\Gamma \tau}e^{j\omega_I \tau}}{\tau}{0}{+\infty} M_s(x)M_I\gpar{x_{-\tau}}\varphi^{b-a}_{\tau,bwd}\gpar{x,p}
\end{equation}

\begin{align}\label{eq:mainLPI_bis}
  \begin{split}
    \gpar{\op{\mathcal{P}}^{\dagger}(\omega_I)}_W[x,p]&=-\frac{j}{\hbar}\sum_b\intg{e^{-\Gamma \tau}e^{-j\omega_I \tau}}{\tau}{0}{+\infty} M_s(x)M_I\gpar{x_{-\tau}}\varphi^{b-a}_{\tau,bwd}\gpar{x,p}\\
    &=\gpar{\op{\mathcal{P}}(\omega_I)}_W^{\star}[x,p]
  \end{split}
\end{align}

Resonance Raman signal intensity:

\begin{align}
  \begin{split}
    I_{Raman}(\omega_s)&=K\frac{2}{1+e^{-\beta \omega}}\omega_I\omega_s^3\frac{1}{2\pi}\frac{1}{\hbar^2}\intg{e^{-j\omega_s t}}{t}{-\infty}{+\infty}\intgd{\frac{\widehat{\rho}_W\gcro{x_0,p_0}}{2\pi\hbar}}{x_0}{p_0}{}\\
    &\times\sum_b M_sM_I\intg{e^{-\Gamma \tau}e^{-j\omega_I\tau}\varphi^{a-b}_{\tau,bwd}(x_0,p_0)}{\tau}{0}{+\infty}\\
    &\times\sum_b M_sM_I\intg{e^{-\Gamma \tau'}e^{j\omega_I\tau'}\varphi^{b-a}_{\tau',bwd}(x_{\tau'},p_{\tau'})}{\tau'}{0}{+\infty}
  \end{split}
\end{align}

The absorption spectrum is obtained from the polarizability operator

\begin{equation}
    \mean{\op{\mathcal{P}}(\omega_I)}=\frac{j}{\hbar}\sum_b \intg{e^{-\Gamma t}e^{j\omega_It}}{t}{0}{+\infty}\intg{\braket{x}{\widehat{\rho}\widehat{M_s}e^{-\frac{j}{\hbar}t\op{H_b}}\widehat{M_I}e^{\frac{j}{\hbar}t\op{H_a}}}{x}}{x}{-\infty}{+\infty}
\end{equation}

which after LPI reads in general as: 

\begin{align}
  \begin{split}
    \mean{\op{\mathcal{P}}(\omega_I)}&=\frac{j}{\hbar}\sum_bM_sM_I\intg{e^{-\Gamma t}e^{j\omega_It}}{t}{0}{+\infty}\intgd{}{x_0}{p_0}{}\\
    &\times\frac{\widehat{\rho}_W\gcro{x_0,p_0}}{2\pi\hbar}\varphi_{t,bwd}^{b-a}(x_0,p_0)
  \end{split}
\end{align}

\section{Wigner transform of a product and symmetrized correlation function}\label{app:wignerprod}

It has been shown by Imre et al.\cite{Imret67} that, being $\Lambda=\overleftarrow{\nabla}\!\!p\,\overrightarrow{\nabla r}-\overleftarrow{\nabla}\!\!r\,\overrightarrow{\nabla p}$, the Wigner transform of a product of operators $\widehat{A}$ and $\widehat{B}$ can be expanded as
\begin{equation}
\begin{split}
        \left(\widehat{A}\widehat{B}\right)_W&=A_W[r,p]e^{\frac{\hbar\Lambda}{2j}}B_W[r,p]\\
        &=A_W[r,p]B_W[r,p]+ A_W[r,p]\frac{\hbar}{2j}\Lambda B_W[r,p], +\mathcal{O}(\hbar^2)
\end{split}
\end{equation}
where in the expansion we  left out terms in $\hbar^2$. By using symmetrized correlation functions, we will cancel the term proportional to $\hbar$. Indeed, for a symmetrized correlation function we have
\begin{align}
  \begin{split}
    C_s(t)
    &=\text{Tr}\left[\widehat{\rho}\widehat{A}\widehat{B}(t)+\widehat{\rho}\widehat{B}(t)\widehat{A}\right]\\
    &=\text{Tr}\left[\widehat{\rho}\widehat{A}\widehat{B}(t)\right]+\text{Tr}\left[\widehat{\rho}\widehat{B}(t)\widehat{A}\right]\\
    &=\text{Tr}\left[\widehat{\rho}\widehat{A}\widehat{B}(t)\right]+\text{Tr}\left[\widehat{A}\widehat{\rho}\widehat{B}(t)\right]\\
    &=\text{Tr}\left[\left(\widehat{\rho}\widehat{A}+\widehat{A}\widehat{\rho}\right)\widehat{B}(t)\right].
  \end{split}
\end{align}
By making use of the expression of the trace and employing Wigner transforms, we obtain
\begin{equation}
    C_s(t)=\iint dx dp \left(\widehat{\rho}\widehat{A}+\widehat{A}\widehat{\rho}\right)_W[r,p]\widehat{B}(t)_W[r,p]
\end{equation}
and $\left(\widehat{\rho}\widehat{A}+\widehat{A}\widehat{\rho}\right)_W[r,p]$ can be expanded as 
\begin{equation}
    \left(\widehat{\rho}\widehat{A}+\widehat{A}\widehat{\rho}\right)_W[r,p]
    =\rho_W[r,p]A_W[r,p] + \mathcal{O}(\hbar^2),
\end{equation}
using the fact that $A_W[r,p]\frac{\hbar}{2j}\Lambda B_W[r,p]= -B_W[r,p]\frac{\hbar}{2j}\Lambda A_W[r,p]$.

One has to make the average:
\begin{equation}
    C_r(t) = \frac{1}{2 \pi \hbar} \iint dx dp \left( \frac{\widehat{\rho}\widehat{A}+ \widehat{A}\widehat{\rho}}{2} \right)_W[x,p] \left(\widehat{B}(t)\right)_W[x,p]
\end{equation}

and finally the Fourier transforms of the symmetrized and the standard correlation functions are related by\cite{Ramirez04,Bonella2005}
\begin{equation}
    I(\omega)=\frac{2}{1+e^{-\beta\hbar\omega}}I_r(\omega),
\end{equation}
which in our case ensures the correct ratio between Stokes and anti-Stokes intensities.


We should note that by calculating the product of the Wigner transforms
instead of the Wigner transform of the product is actually a better 
approximation of the symmetrized correlation function (here up to second
order in $\hbar$) than of the initially desired normal one (at first order in $\hbar$). 
It could also be noted that these two functions have the same symmetry under time reversal.

\bibliography{ref}

\end{document}

%% file: article.preamble.tex
\newcommand{\gpar}[1]{\left(#1\right)}

\newcommand{\gcro}[1]{\left[#1\right]}

\newcommand{\braket}[3]{\left<#1\middle|#2\middle|#3\right>}
\newcommand{\mean}[1]{\left<#1\right>}

\newcommand{\op}[1]{\widehat{#1}}

\newcommand{\intg}[4]{\int_{#3}^{#4} \mathrm{d}#2 \, #1}
\newcommand{\intgd}[4]{\int\!\!\!\!\int_{#4} \mathrm{d}#2 \, \mathrm{d}#3 \, #1}
\newcommand{\intgm}[4]{\int\!\!\cdots\!\!\int_{#4} \mathrm{d}#2 \cdots \mathrm{d}#3 \, #1}

\newcommand{\botacc}[2]{\underbrace{#1}_{\let\scriptstyle\textstyle\substack{#2}}}